\documentclass[11pt]{article}

\usepackage[latin1]{inputenc}

\usepackage{amstext,amsmath,amssymb,amsfonts,bbm,amsmath}
\usepackage{extarrows}
\usepackage{xcolor} 
\usepackage{hyperref}
\hypersetup{
    colorlinks=false,
    menubordercolor=red,
    linkbordercolor=blue
}
\usepackage{authblk}
\usepackage{caption} 
\usepackage{epsfig} 
\usepackage{color} 
\usepackage{graphicx} 
\usepackage{braket} 
\usepackage{slashed} 
\usepackage{dsfont} 
\usepackage{amsthm}

\usepackage{cite}

\allowdisplaybreaks[4]

\usepackage{psfrag}
\usepackage{tikz} 
\usetikzlibrary{arrows}
\usetikzlibrary{plotmarks}
\usetikzlibrary{external}
\usetikzlibrary{patterns}
\usepackage{pgfplots}
\pgfplotsset{compat=1.9}
\usetikzlibrary{patterns}
\usetikzlibrary{calc}
\usetikzlibrary{automata,positioning}

\usepackage{float}  
\usepackage{subfig}
\usepackage[lmargin=50pt,rmargin=60pt,tmargin=60pt,bmargin=65pt]{geometry}

\newcommand{\be}{\begin{equation}}
\newcommand{\ee}{\end{equation}} 

\newcommand{\f}{\frac}
\newcommand{\p}{\partial}

\let\a=\alpha \let\b=\beta      
\let\z=\zeta

\let\G=\Gamma     \let\X=F

\newcommand{\cN}{\mathcal{N}}
\newcommand{\cO}{\mathcal{O}}

\newcommand{\gt}{\tilde{g}}
\newcommand{\rt}{\tilde{r}}

\DeclareMathOperator{\im}{\mathrm{i}}

\newcommand{\mba}{\mathbf{a}}
\newcommand{\mbb}{\mathbf{b}}
\newcommand{\mbc}{\mathbf{c}}
\newcommand{\mbd}{\mathbf{d}}
\newcommand{\mbe}{\mathbf{e}}
\newcommand{\mbf}{\mathbf{f}}
\newcommand{\mbg}{\mathbf{g}}
\newcommand{\mbh}{\mathbf{h}}

\allowdisplaybreaks[4]

\numberwithin{equation}{section}

\theoremstyle{remark}

\begin{document}

\title{\bf The tri-fundamental quartic model}

\author[1]{Dario Benedetti}
\author[1,2,3]{Razvan Gurau}
\author[1]{Sabine Harribey}

\affil[1]{\normalsize \it 
 CPHT, CNRS, Ecole Polytechnique, Institut Polytechnique de Paris, Route de Saclay, \authorcr 91128 PALAISEAU, 
 France
 \authorcr
emails: dario.benedetti@polytechnique.edu, rgurau@cpht.polytechnique.fr, sabine.harribey@polytechnique.edu
 \authorcr \hfill }

\affil[2]{\normalsize\it 
Perimeter Institute for Theoretical Physics, 31 Caroline St. N, N2L 2Y5, Waterloo, ON,
Canada
 \authorcr \hfill}

\affil[3]{\normalsize\it 
Heidelberg University, Institut f\"ur Theoretische Physik, Philosophenweg 19, 69120 Heidelberg, Germany
 \authorcr \hfill}
 
\date{}

\maketitle

\hrule\bigskip

\begin{abstract}
We consider a multi-scalar field theory with either short-range or long-range free action and with quartic interactions that are invariant under $O(N_1)\times O(N_2) \times O(N_3)$ transformations, of which the scalar fields form a tri-fundamental representation.
We study the renormalization group fixed points at two loops at finite $N$ and in various large-$N$ scaling limits for small $\epsilon$, the latter being either the deviation from the critical dimension or from the critical scaling of the free propagator. In particular, for the homogeneous case $N_i = N$ for $i=1,2,3$, we study the subleading corrections to previously known fixed points.
In the short-range model, for $\epsilon N^2\gg 1$, we find complex fixed points with non-zero tetrahedral coupling, that at leading order reproduce the results of Giombi et al.\ \cite{Giombi:2017dtl}; the main novelty at next-to-leading order is that the critical exponents acquire a real part, thus allowing a correct identification of some fixed points as IR stable.
In the long-range model, for $\epsilon N \ll 1 $, we find again complex fixed points with non-zero tetrahedral coupling, that at leading order reproduce the line of stable fixed points of Benedetti et al. \cite{Benedetti:2019eyl}; at next-to-leading order, this is reduced to a discrete set of stable fixed points. One difference between the short-range and long-range cases is that, in the former the critical exponents are purely imaginary at leading-order and gain a real part at next-to-leading order, while for the latter the situation is reversed.
\end{abstract}

\hrule\bigskip

\tableofcontents

\section{Introduction}
\label{sec:introduction}

Multi-scalar models with quartic interactions are a broad class of field theories including some extensively studied models, such as the $O(\cN)$ model.  
The latter describes some of the most important universality classes, such as the Ising and Heisenberg models, but other multi-scalar models, with smaller symmetry groups, are also of general interest (see for example \cite{Pelissetto:2000ek,Kleinert:2001ax} and references therein). In fact, being able to classify or better understand all the possible universality classes appearing in such models would be of great theoretical appeal. Efforts in this direction have been made for example in \cite{Brezin:1973jt,Michel:1983in,Michel:1985,Toledano:1985,Hatch:1985,Vicari:2006xr,Osborn:2017ucf,Rychkov:2018vya,Codello:2018nbe,Codello:2020lta,Hogervorst:2020gtc,Osborn:2020cnf}, but clearly a full classification becomes daunting as the number $\cN$ of fields increases.
It is then natural to try to broaden our understanding by gradually breaking the maximal symmetry group, i.e.\ the $O(\cN)$ group, to smaller ones, which of course can be done in many ways.
One much studied case is the model with symmetry $O(N_1)\times O(N_2)$, with $N_1 N_2 =\cN$, which was named \emph{bi-fundamental model} recently in \cite{Rychkov:2018vya}, but which has a long history (e.g.\ \cite{Kawamura:1988,Kawamura:1990,Pelissetto:2001fi,Gracey:2002pm,Delamotte:2003dw,Kompaniets:2020,Henriksson:2020fqi}).

In this paper we go one step further in the same direction, and consider a \emph{tri-fundamental model}, with symmetry group $O(N_1)\times O(N_2) \times O(N_3)$, and $N_1 N_2 N_3=\cN$.
Whereas the $O(\cN)$ model has a single coupling, and the bi-fundamental model has two, the tri-fundamental model has five independent couplings (with the corresponding interactions being known as tetrahedron, double trace, and pillows, the latter being of three different types), making its system of beta functions more involved. For this reason, we study its fixed points either numerically, for specific values of the $N_i$'s, or in some large-$N$ scaling limits, with either one, two, or all three  of the $N_i$'s being taken to infinity.
In the homogeneous case $N_i = N$, for $i=1,2,3$, the model reduces to the $O(N)^3$ \emph{tensor model}, which has already been studied in the strict large-$N$ limit \cite{Giombi:2017dtl,Benedetti:2019eyl}.

Tensor models are particularly interesting because at large $N$ they are dominated by melonic diagrams \cite{Bonzom:2011zz,RTM,Carrozza:2015adg}. The melonic limit is different from both the vector and matrix large $N$ limits \cite{Brezin:1994eb}: it is richer than the large $N$ limit of vectors but is more manageable than the planar limit of matrices.
Consequently, at large $N$, renormalization group fixed points of tensor models in $d$ dimensions give rise to a new family of conformal field theories (CFTs) which are analytically accessible \cite{Klebanov:2016xxf,Giombi:2017dtl,Prakash:2017hwq,Benedetti:2017fmp,Gubser:2018yec,Giombi:2018qgp,Benedetti:2018ghn,Popov:2019nja,Benedetti:2019rja} (see also \cite{Delporte:2018iyf,Klebanov:2018fzb,Gurau:2019qag,Benedetti:2020seh} for reviews and more references). We call this new family of CFTs melonic.
The $O(N)^3$ bosonic tensor model with quartic interactions is one of the simplest tensor models we can study. This model is also known as the CTKT model as it was introduced in zero dimension by Carozza and Tanasa in \cite{Carrozza:2015adg} and generalized to $d$ dimensions by Klebanov and Tarnopolski in \cite{Klebanov:2016xxf}. It was studied further in \cite{Giombi:2017dtl}, where in $d=4-\epsilon$ dimensions and  in the melonic limit it was found to have non-trivial fixed points, which however correspond to complex CFTs. Moreover, the critical exponents determining the approach to the fixed points are purely imaginary, and the trajectories around the fixed points form concentric cycles, never really reaching them; in the language of dynamical systems, the fixed point is a center equilibrium. A similar model was then studied  in \cite{Benedetti:2019eyl}, with the same symmetry and interactions, but with a long-range kinetic term. A line of infrared stable fixed points  was found, parametrized by a purely imaginary and exactly marginal tetrahedron coupling. Surprisingly, the resulting large-$N$ CFT, which was studied in \cite{Benedetti:2019ikb,Benedetti:2020yvb},  appears to be unitarity, despite the fact that the tetrahedral coupling is imaginary. 
 
These results are valid in the large $N$ limit, thus they leave open several questions. How do the subleading corrections in $1/N$ change them? In particular, generalizing to a $O(N_1)\times O(N_2)\times O(N_3)$ symmetry in the short-range case, can we find at small $N_i$ or in some scaling limit real stable fixed points with non-zero tetrahedral coupling? 
For the long-range case, what becomes of the line of fixed points at next-to-leading order? Do we have a breaking of unitarity at subleading orders, for example signaled by the critical exponents having complex $1/N$ corrections?

The study of the $1/N$ corrections to the melonic limit turns out to be surprisingly involved. This is due to the fact that the 
tetrahedron coupling receives no radiative corrections at large $N$, and therefore its beta function is either trivial (long-range case) or determined solely by the wave-function renormalization (short-range case), the latter only starting with a (two-loop) cubic term.
At order $1/N$, the beta function of the tetrahedral coupling acquires  a (one-loop) quadratic term, destroying its exact marginality in the long-range model, and creating in the short-range model a delicate competition with the cubic term, the latter being leading in $1/N$ but subleading in the coupling.
In order to disentangle the effects of this quadratic term one needs to analyze scaling regimes defining a hierarchy between $1/N$ and $\epsilon$, where $\epsilon$ is either defined as the deviation from the critical dimension in the short-range case, i.e.\ $\epsilon=4-d$, or as the deviation from the critical scaling of the propagator in the long-range case, i.e.\  $C(p)=1/p^{(d+\epsilon)/2}$. We carry out this analysis below.

\paragraph{Plan of the paper and summary of results.}
In this paper we study the tri-fundamental model $O(N_1)\times O(N_2)\times O(N_3)$ with quartic interactions, both in the short-range and long-range  versions. Their finite-$N$ beta functions can be obtained as particular cases of general multi-scalar models, for which we use standard results such as \cite{ZinnJustin:2002ru} for the short-range case, and the three-loop results recently obtained in \cite{Benedetti:2020rrq} for the long-range case.
We are interested in fixed points with no enhanced symmetry such as $O(N_1 N_2 N_3)$ or $O(N_1)\times O(N_2 N_3)$, and therefore we are in particular interested in fixed points with non-vanishing tetrahedral coupling. In fact, the latter is the single coupling which is most characteristic of the full symmetry group, being capable alone to generate all the others by RG flow. Moreover, it is the coupling that in the $O(N)^3$ model leads to a melonic dominance at large $N$.

In section \ref{sec:short-range}, after a quick review of the short-range multi-scalar model, we compute the beta functions and fixed points of the short-range $O(N_1)\times O(N_2)\times O(N_3)$ model at two loops. First, in section \ref{sec:short-range_num}, we look for numerical solutions of the fixed point equations at finite $N_i$, and we find that there is \emph{no} real fixed point with non-vanishing tetrahedral coupling that is stable in all five directions in the range $2 \le N_i \le 50$. Then, in sections \ref{sec:short-range_vector} and \ref{sec:short-range_matrix}, we compute the fixed points respectively in the vector-like ($N_1 \rightarrow \infty$; $N_2$ and $N_3$ fixed) and in the matrix-like ($N_2= cN_1= N \rightarrow \infty$; $c$ and $N_3$ fixed) limits. Like at finite $N_i$, in both cases we conclude that there is no real stable fixed point with non-vanishing tetrahedral coupling; however, in the matrix-like case we find a complex stable fixed point. Finally, in section \ref{sec:short-range_triple}, we study the large $N$ limit and its first subleading corrections in the case $N_1=N_2=N_3=N$, and with a single coupling for the three pillow interactions, corresponding to the $O(N)^3$ tensor model. At leading order our results agree with those of \cite{Giombi:2017dtl}. It turns out that, due to the quadratic term at order $1/N$ in the beta function of the tetrahedral coupling, to which we alluded before, in order to study the $1/N$ corrections to the leading-order fixed point we must consider $\epsilon N^2 \gg 1$. In this regime we find a fixed point for which all three critical exponents have positive real parts. The real part is $\cO(1)$ for the tetrahedral coupling and $\cO(1/N)$ for the other two.

Next, in section \ref{sec:long-range}, we study the long-range case. After a quick review of the long-range multi-scalar model,  in section \ref{sec:long-range_equalN} we directly set $N_1=N_2=N_3=N$ to study the bosonic $O(N)^3$ tensor model, with a single coupling for the three pillow interactions. We study the fixed points and critical exponents at two loops, up to and including order $1/N$. 
At leading order, we reproduce the results of \cite{Benedetti:2019eyl}, that is, at $\epsilon=0$, i.e.\ for a propagator such that the quartic interactions have dimension $d$, we find a line of stable real fixed points for the pillow and double-trace couplings parametrized by an exactly marginal tetrahedral coupling; we stress that the latter needs to be taken purely imaginary for the other fixed points and their critical exponents to be real. 
At order $1/N$, due to the occurrence of non-vanishing quadratic terms in its beta function, the tetrahedral coupling is no longer exactly marginal, and  in order to find non-trivial but perturbatively accessible fixed points, it turns out that we must turn on $\epsilon$ and we must consider the regime $ \epsilon N \ll 1$.\footnote{This is somewhat similar to what has been observed by Fleming et al.\ in \cite{Fleming:2020qqx} for the  $O(N)$ model with $(\phi^2)^3$ interaction in $d=3-\epsilon$ dimensions. At $\epsilon=0$, the coupling of $(\phi^2)^3$ has a vanishing beta function at large-$N$, thus leading to a line of fixed points; Fleming et al.\ pointed out that in order to find a precursor of such line of fixed points at subleading orders in $1/N$, one needs to turn on $\epsilon$ and look in the $(d,N)$-plane for lines of fixed points parametrized by $\a=\epsilon N$. Of course, at fixed $\a$  we do not have a line anymore, but isolated fixed points. We moreover take $ \epsilon N \ll 1$ as we wish to rely on perturbation theory, rather than on functional renormalization group methods, as done instead in \cite{Fleming:2020qqx}.}
In this regime, we find a purely imaginary fixed point for the tetrahedral coupling; its value being imaginary, the reality of the leading critical exponents is not spoiled. However, at order $1/N$ we find purely imaginary corrections for the pillow and double-trace critical exponents; at the same order, the tetrahedron critical exponent is real, but it also acquires an imaginary part at order $1/N^{3/2}$. 

We thus see two similar situations. In the short-range model, the leading-order fixed point is real for the tetrahedral coupling and purely imaginary for the pillow and double-trace couplings. The critical exponent of the tetrahedral coupling is real while the others are purely imaginary. The $1/N$ corrections bring real parts to the 
pillow and double-trace critical exponents (and a small imaginary part to the tetrahedral critical exponent). These real parts lead to a stable fixed point.

In the long-range model, we have the reverse: at leading order the stable line of fixed points corresponds to purely imaginary tetrahedral coupling and real pillow and the double-trace couplings. The critical exponents of the pillow and double trace are real (the one of the tetrahedral coupling is of course zero). At higher orders in $1/N$ the fixed line collapses to isolated fixed points. The pillow and double-trace critical exponents acquire small imaginary parts, and the tetrahedron critical exponent is real at the first non-trivial order, and complex beyond that. The fixed point is  stable in all three directions, but the critical exponents have non zero imaginary parts. The unitarity of the large-$N$ melonic CFT is broken by the $1/N$ corrections.

\bigskip
\bigskip

\section{The short-range tri-fundamental model}
\label{sec:short-range}

\subsection{The short-range multi-scalar model}
\label{sec:short-range_ms}
The short-range multi-scalar model with quartic interactions in dimension $d$ is defined by the action:
\begin{align}
		S[\phi]  \, &= \, \int d^dx \, \bigg[ \frac{1}{2} \partial_{\mu} \phi_\mba(x) \partial_{\mu} \phi_{\mba}(x)
		\, + \, \frac{1}{4!} \, \lambda_{\mba \mbb \mbc \mbd} \phi_{\mba}(x) \phi_{\mbb}(x) \phi_{\mbc}(x) \phi_{\mbd}(x) \bigg] \, ,
	\end{align}
where the indices take values from 1 to $\cN$, and a summation over repeated indices is implicit.
For the Euclidean theory in $d=4-\epsilon$ dimension the beta function up to two loops in the minimal subtraction scheme \cite{ZinnJustin:2002ru} is:
\begin{align} \label{eq:beta-general}
		\beta_{\mba \mbb \mbc \mbd} \, &= \, - \, \epsilon \gt_{\mba \mbb \mbc \mbd}
		\, + \, \left(\gt_{\mba \mbb \mbe \mbf}\gt_{\mbe \mbf \mbc \mbd} + 2 \textrm{ terms} \right) 
		\, - \, \left(\gt_{\mba \mbb \mbe \mbf}\gt_{\mbe \mbg \mbh \mbc}\gt_{\mbf \mbg \mbh \mbd}+ 5 \textrm{ terms} \right) \nonumber\\
		&\quad + \, \frac{1}{12} \left(\gt_{\mba \mbb \mbc \mbe}\gt_{\mbe \mbf \mbg \mbh}\gt_{\mbf \mbg \mbh \mbd} + 3 \textrm{ terms} \right) +\mathcal{O}(\gt^4) \,,
	\end{align}
where we rescaled the renormalized coupling to $\gt_{\mba \mbb \mbc \mbd} = g_{\mba \mbb \mbc \mbd} (4\pi)^{-d/2}/\Gamma(d/2)$.

By imposing various symmetry restrictions on the interaction one obtains different models which have been extensively studied (see for example \cite{Pelissetto:2000ek,Kleinert:2001ax,Vicari:2006xr,Osborn:2017ucf,Rychkov:2018vya} and references therein). We study here the case with $O(N_1)\times O(N_2) \times O(N_3)$ invariance, which is relatively new.

\subsection{The short-range  tri-fundamental model}
The fields in the tri-fundamental model are rank-3 tensor fields transforming in the tri-fundamental representation of $O(N_1)\times O(N_2) \times O(N_3)$. This is made manifest by writing the index $\mba$ as a triplet $\mba=(a_1,a_2,a_3)$, where the first, second and third index correspond to the $O(N_1)$, $O(N_2)$ and $O(N_3)$ group respectively:
\be
\phi_{a_1 a_2 a_3} \to \sum_{b_1 b_2 b_3}^{1\ldots N} R^{(1)}_{a_1 b_1}R^{(2)}_{a_2 b_2}R^{(3)}_{a_3 b_3}\phi_{b_1 b_2 b_3}\,, \;\;\;\; R^{(i)}\in O(N_i)\,.
\ee
Notice that since each orthogonal group contains a $\mathbb{Z}_2$ subgroup, in order to have a faithful action of the symmetry group, we should quotient $O(N_1)\times O(N_2) \times O(N_3)$ by a $\mathbb{Z}_2^3$, which acts trivially. As this is irrelevant to our study, we will stick to the  unquotiented version of the symmetry group.

\begin{figure}[ht]
\begin{center}
\includegraphics[width=0.5\textwidth]{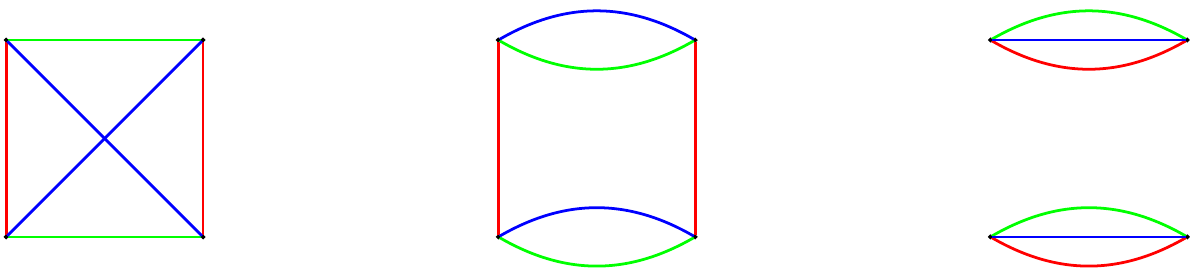}
 \caption{Graphical representation of the quartic $O(N_1)\times O(N_2) \times O(N_3)$ invariants: each vertex represents a tensor field, and each edge represents a Kronecker delta contracting two indices, appropriately color-coded to distinguish the three indices of a tensor. From left to right: the tetrahedron, the pillow, and the double-trace (there are three pillow contractions, distinguished by the color of the vertical edge).} \label{fig:interactions}
 \end{center}
\end{figure}

Under such symmetry transformation, the most general invariant tensor structure for the coupling is\footnote{The normalization has be chosen so that the couplings are normalized by $1/4$ and not $1/4!$, as usually done in tensor models.}:
	\begin{align} \label{eq:coupling-trifund}
		\gt_{\mba \mbb \mbc \mbd} \, &= \, \gt \left(\delta^t_{\mba \mbb \mbc \mbd} + 5 \textrm{ terms} \right)
		\, + \, \sum_{i=1,2,3} \gt_{p,i} \left(\delta^{p,i}_{\mba \mbb ;\mbc \mbd} + 5 \textrm{ terms} \right)
		\, + \, 2\gt_d\left(\delta^d_{\mba \mbb \mbc \mbd} + 2 \textrm{terms} \right) \, ,
	\end{align}
where: 
	\begin{gather}
		\delta_{\mba \mbb} =\prod_{i=1}^3\delta_{a_i b_i}\, , \qquad \delta^t_{\mba \mbb\mbc\mbd} \, = \, \delta_{a_1 b_1}  \delta_{c_1 d_1} \delta_{a_2 c_2}  \delta_{b_2 d_2 } \delta_{a_3 d_3}   \delta_{b_3 c_3} \, , \nonumber\\[4pt]
		\delta^{p, i}_{\mba\mbb; \mbc\mbd } \, = \, \delta_{a_ic_i} \delta_{b_id_i} \prod_{j\neq i}  \delta_{a_jb_j}  \delta_{c_jd_j} \, ,
		\qquad  \delta^d_{\mba\mbb; \mbc\mbd }  = \delta_{\mba \mbb}  \delta_{\mbc \mbd}  \, .
		\label{eq:deltas-nohat}
	\end{gather}
The labels $t$, $p$, and $d$ stand for tetrahedron, pillow, and double-trace, respectively. The first two names describe the graphical representation of the corresponding invariants \cite{Carrozza:2015adg} (see Fig.~\ref{fig:interactions}), while the third one, which is the square of the unique quadratic invariant, is named in analogy to matrix models \cite{Giombi:2017dtl}. 

Substituting \eqref{eq:coupling-trifund} in equation \eqref{eq:beta-general} and truncating at one loop, we obtain the beta functions:
	\begin{align}
		\beta_t=&-\epsilon\tilde{g}+4\Big[6\tilde{g}\tilde{g}_d+2(\tilde{g}_{p,1}\tilde{g}_{p,2}+\tilde{g}_{p,1}\tilde{g}_{p,3}+\tilde{g}_{p,2}\tilde{g}_{p,3})
		+\tilde{g}((1+N_1)\tilde{g}_{p,1}+(1+N_2)\tilde{g}_{p,2}+(1+N_3)\tilde{g}_{p,3})\Big] \, , \nonumber\\
		\beta_{p,i}=&-\epsilon\tilde{g}_{p,i}+2\Big[12\tilde{g}_d\tilde{g}_{p,i}+4\tilde{g}(\tilde{g}_{p,i+1}+\tilde{g}_{p,i+2})
		+4\tilde{g}_{p,i+1}\tilde{g}_{p,i+2}+(2+N_i)\tilde{g}^2 \crcr
		& +2\tilde{g}_{p,i}((1+N_{i+1})\tilde{g}_{p,i+1}+(1+N_{i+2})\tilde{g}_{p,i+2}+(N_{i+1}+N_{i+2})\tilde{g})+\tilde{g}_{p,i}^2(4+N_i+N_{i+1}N_{i+2})\Big] \nonumber\\
		\beta_d=&-\epsilon\tilde{g}_d+2\Big[\tilde{g}_d^2(8+N_1N_2N_3)+3(\tilde{g}_{p,1}^2+\tilde{g}_{p,2}^2+\tilde{g}_{p,3}^2)
		+2\tilde{g}(\tilde{g}_{p,1}+\tilde{g}_{p,2}+\tilde{g}_{p,3})\crcr
		& +2 (N_1\tilde{g}_{p,2}\tilde{g}_{p,3}+N_2\tilde{g}_{p,1}\tilde{g}_{p,3}+N_3\tilde{g}_{p,1}\tilde{g}_{p,2}) +2\tilde{g}_d\tilde{g}(N_1+N_2+N_3) \nonumber\\
		&+2\tilde{g}_d((1+N_1+N_2N_3)\tilde{g}_{p,1}+(1+N_2+N_1N_3)\tilde{g}_{p,2}+(1+N_3+N_1N_2)\tilde{g}_{p,3}) \Big] \, ,
	\label{eq:beta^(4)}
	\end{align}
where $i\in\{1,2,3\}\,\text{mod}\, 3$, i.e.\ $g_{p,4}=g_{p,1}$ and $g_{p,5}=g_{p,2}$.
The two-loop terms can be obtained by computer algebra but they are too long to write here and we will only use them in section~\ref{sec:short-range_triple}.
In Appendix~\ref{app:grad_flow} we write the system \eqref{eq:beta^(4)} as a gradient flow.

Notice that the model with only double-trace interaction has an enhanced symmetry, being invariant under field transformations in the fundamental representation of $O(N_1 N_2 N_3)$; that is, it is the usual $O(\cN)$ model in disguise.
Similarly, if all the couplings except the double-trace and one pillow, e.g.\ $\gt_{p,1}$, are zero, then the model has the symmetry group  $O(N_1)\times O(N_2 N_3)$, and it is a bi-fundamental model in disguise. 
Such symmetry enhancements are reflected in the fact that the couplings set to zero are not turned on by the renormalization group flow.
Keeping instead at least two pillows, or just the tetrahedron, will break the symmetry back to $O(N_1)\times O(N_2) \times O(N_3)$, and the flow will generate the remaining couplings.
The tetrahedron is in fact the single coupling which is most characteristic of the full symmetry group, being capable alone to generate all the others by RG flow. Moreover, it is the coupling that in the $O(N)^3$ model leads to a melonic dominance at large $N$.
Therefore, in most of the following we will only be interested in fixed points with non-vanishing tetrahedron coupling, $\tilde{g}\ne 0$.

\paragraph{Remark.}

Let us consider the case of real coupling constants.
The tetrahedron interaction is not positive definite, thus the most general potential can be unstable. However, for particular choices of (real) couplings, the pillow and double-trace can dominate over the negative direction of the tetrahedron, thus making the whole interaction positive. In particular, as shown in  \cite{Michel:1983in} (see also \cite{ZinnJustin:2007zz}), based on the gradient flow representation \cite{Wallace:1974dy}, at order $\epsilon$ any non-trivial fixed point corresponds to a positive interaction. Moreover, we also know that an unstable interaction cannot flow to a stable one \cite{Rychkov:2018vya}.
As a consequence, any non-trivial fixed point with $\tilde{g}\neq 0$ must also have at least some of the other couplings non-zero, and so must also any initial condition that flows to such a fixed point.

\subsection{Numerical solutions for small $N_i$}
\label{sec:short-range_num}

Even at the one-loop level it is hard to solve the beta functions (\ref{eq:beta^(4)}) for generic values of $N_i$.
In this subsection we solve numerically for fixed points at low $N_i$.
In the following, we only search for fixed points at one-loop with $\tilde{g}\ne 0$ and real critical couplings.

We define the critical coupling vector by
	\begin{align}
		\vec{g}_\star \, \equiv \, \big( \tilde{g}^\star, \, \tilde{g}_{p,1}^\star, \, \tilde{g}_{p,2}^\star, \, \tilde{g}_{p,3}^\star, \, \tilde{g}_d^\star \big) \, .
	\end{align}
The stability matrix of the fixed point is given by 
	\begin{align}
		M_{ab} \, = \, \frac{\partial \beta_a(\vec{g})}{\partial g_b} \bigg|_{\vec{g}_\star} \, ,
	\end{align}
where $a, b=\{t, p1, p2, p3, d\}$.
We arrange the eigenvalues of this matrix in a vector denoted by $\vec{\omega}$.
If the eigenvalue $\omega_a$ is positive, then the fixed point is stable in the corresponding eigendirection.

\medskip

We have numerically checked that there is \emph{no} real fixed point with $\tilde{g}\ne 0$ that is stable in all five directions in the range $2 \le N_i \le 50$.
We explicitly show some examples below.

\subsubsection{$N_1=2$, $N_2=2$}
By fixing $N_1=N_2=2$, there is no fixed point in the range of $2 \le N_3 \le21$.
At $N_3 = 22$, we find four fixed points:   
	\begin{align}
		\vec{g}_\star \, &= \, \left( \frac{\epsilon}{1200}, \, -\frac{\epsilon}{600}, \, -\frac{\epsilon}{600}, \, \frac{\epsilon}{400}, \, \frac{\epsilon}{800} \right) \, , \qquad
		\vec{\omega} \, = \, \big( -6.72\epsilon, \, 6\epsilon, \, -4.8\epsilon, \, 0.48\epsilon, \, -0.48\epsilon \big) \, , \\
		\vec{g}_\star \, &= \, \left( \frac{7\epsilon}{8304}, \, -\frac{7\epsilon}{4152}, \, -\frac{7\epsilon}{4152}, \, \frac{7\epsilon}{2768}, \, \frac{59\epsilon}{49824} \right) \, ,\qquad
		\vec{\omega} \, = \, \big( -6.79\epsilon, \, 6\epsilon, \, -4.85\epsilon, \, 0.485\epsilon, \, -0.485\epsilon \big) \, , \\
		\vec{g}_\star \, &= \, \left( \frac{7\epsilon}{6912}, \, -\frac{35\epsilon}{20736}, \, -\frac{35\epsilon}{20736}, \, \frac{49\epsilon}{20736}, \, \frac{157\epsilon}{124416} \right) \, ,\qquad
		\vec{\omega} \, = \, \big( -6.70\epsilon, \, 6\epsilon, \, -4.76\epsilon, \, 0.486\epsilon, \, -0.486\epsilon \big) \, , \\
		\vec{g}_\star \, &= \, \left( \frac{\epsilon}{976}, \, -\frac{5\epsilon}{2928}, \, -\frac{5\epsilon}{2928}, \, \frac{7\epsilon}{2928}, \, \frac{7\epsilon}{5856} \right) \, , \qquad
		\vec{\omega} \, = \, \big( -6.78\epsilon, \, 6\epsilon, \, -4.82\epsilon, \, -0.491\epsilon, \, -0.491\epsilon \big) \, .
	\end{align}
There is no fixed point which is stable in all five directions.
In the range of $23 \le N_3 \le33$, we find similar four types of fixed points.
At $N_3 = 34$, we find six fixed points:
	\begin{align}
		\vec{g}_\star \, &= \, \left( \frac{\epsilon}{1008}, \, -\frac{\epsilon}{1008}, \, 0, \, \frac{\epsilon}{756}, \, \frac{\epsilon}{6048} \right) \, , \qquad
		\vec{\omega} \, = \, \big( 6\epsilon, \, 5.58\epsilon, \, -4\epsilon, \, -3.86\epsilon, \, 0 \big) \, , \\
		\vec{g}_\star \, &= \, \left( \frac{11(3445\mp203\sqrt{97})\epsilon}{62966016}, \, -\frac{11(-6479\pm\sqrt{97})\epsilon}{62966016}, \, -\frac{11(-6479\pm\sqrt{97})\epsilon}{62966016}, \,
		\frac{11(3171\pm67\sqrt{97})\epsilon}{20988672}, \right. \nonumber\\
		& \left. \qquad \, \frac{(314123\mp6325\sqrt{97})\epsilon}{377796096} \right) \, ,\nonumber\\
		&\quad \vec{\omega} \, = \, \big( -6.69\epsilon, \, 6\epsilon, \, -5.39\epsilon, \, 3.21\epsilon, \, -3.21\epsilon \big) \qquad ({\rm for\ upper}) \, , \\
		&\quad \vec{\omega} \, = \, \big( -6.30\epsilon, \, 6\epsilon, \, -5.00\epsilon, \, 3.21\epsilon, \, -3.21\epsilon \big) \qquad ({\rm for\ lower}) \, , \\
		\vec{g}_\star \, &= \, \left( \frac{(161-10\sqrt{97})\epsilon}{259536}, \, \frac{(-589+3\sqrt{97})\epsilon}{519072}, \, \frac{(-589+3\sqrt{97})\epsilon}{519072}, \,
		\frac{(428+7\sqrt{97})\epsilon}{259536}, \, \frac{(428+7\sqrt{97})\epsilon}{519072} \right) \, ,\nonumber\\
		&\quad \vec{\omega} \, = \, \big( -6.37\epsilon, \, 6\epsilon, \, -5.13\epsilon, \, 3.06\epsilon, \, -3.06\epsilon \big) \qquad ({\rm for\ upper}) \, , \\
		&\quad \vec{\omega} \, = \, \big( -6.63\epsilon, \, 6\epsilon, \, -5.25\epsilon, \, 3.38\epsilon, \, -3.38\epsilon \big) \qquad ({\rm for\ lower}) \, .
	\end{align}
The other solution is given by the first solution with exchanging $\bar{g}_{p,1}^\star$ and $\bar{g}_{p,2}^\star$ with the same eigenvalue vector.
There is no fixed point which is stable in all five directions.

\subsubsection{$N_1=2$, $N_2=3$}
By fixing $N_1=2$ and $N_2=3$, there is no fixed point in the range of $2 \le N_3 \le45$.
At $N_3 = 46$, we find two fixed points:
	\begin{align}
		\vec{g}_\star \, &= \, \left( \frac{17(126955\pm3\sqrt{1345})\epsilon}{2844275280}, \, 0, \, - \frac{17(126955\pm3\sqrt{1345})\epsilon}{2844275280}, \,
		\frac{(126955\pm3\sqrt{1345})\epsilon}{129285240}, \right. \nonumber\\
		& \left. \qquad \, \frac{(243160\mp461\sqrt{1345})\epsilon}{2844275280} \right) \, ,\nonumber\\
		&\quad \vec{\omega} \, = \, \big( 6\epsilon, \, 5.67\epsilon, \, -4.27\epsilon, \, -4.17\epsilon, \, -0.24\epsilon \big) \qquad ({\rm for\ upper}) \, , \\
		&\quad \vec{\omega} \, = \, \big( 6\epsilon, \, 5.66\epsilon, \, -4.26\epsilon, \, -4.16\epsilon, \, 0.24\epsilon \big) \qquad ({\rm for\ lower}) \, .
	\end{align}
There is no fixed point which is stable in all five directions.
At least up to $N_3=1000$, we find the same type of two fixed points, and they are unstable in some of the directions.

\subsection{Vector-like limit}
\label{sec:short-range_vector}

We now consider the limit $N_1 \rightarrow \infty$ while keeping $N_2$ and $N_3$ fixed. 
We define the new couplings:
\begin{align}
\tilde{g}_S=\tilde{g}+\tilde{g}_{p,1} \,, \;\;\; \tilde{g}_D=\tilde{g}-\tilde{g}_{p,1} \,, \;\;\; \tilde{g}_2=\tilde{g}_d+\frac{\tilde{g}_{p,2}}{N_2}+\frac{\tilde{g}_{p,3}}{N_3} \,,
\end{align}
which correspond to orthogonal operators at large $N_1$, and thus their beta functions will decouple. We furthermore rescale the couplings in order to obtain a large-$N_1$ expansion:
\begin{equation}
\tilde{g}_S=\frac{\bar{g}_S}{N_1} \,, \;\;\; \tilde{g}_D=\frac{\bar{g}_D}{N_1} \,, \;\;\;\tilde{g}_{p,i}=\frac{\bar{g}_{p,i}}{N_1} \,, \;\;\; \tilde{g}_2=\frac{\bar{g}_2}{N_1} \,.
\end{equation}
The three loop terms are suppressed in $1/N_1$ and at leading order we obtain the following beta functions:
\begin{align}
\beta_S&=-\epsilon\bar{g}_S+2\bar{g}_S^2\crcr
\beta_{D}&=-\epsilon\bar{g}_D-2\bar{g}_D^2\crcr\crcr
\beta_{p,2}&=-\epsilon\bar{g}_{p,2}+4\bar{g}_S\bar{g}_{p,2}+2N_3\bar{g}_{p,2}^2\crcr
\beta_{p,3}&=-\epsilon\bar{g}_{p,3}+4\bar{g}_S\bar{g}_{p,3}+2N_2\bar{g}_{p,3}^2\crcr
\beta_2&=-\epsilon\bar{g}_2 +4\bar{g}_S\bar{g}_2+2N_2N_3\bar{g}_2^2\, . 
\label{eq:beta_largeN1}
\end{align}

We can then solve for fixed points. We obtain the following 32 fixed points:
\begin{align}
\bar{g}_S^{\star}&=\{0,\frac{\epsilon}{2}\} \, , \;\; \bar{g}_D^{\star}=\{0,-\frac{\epsilon}{2}\}\, , \crcr
\bar{g}_{p,2}&=\{0,\pm \frac{\epsilon}{2N_3}\} \, , \;\; \bar{g}_{p,3}=\{0,\pm \frac{\epsilon}{2N_2}\} \, , \;\; \bar{g}_{2}=\{0,\pm \frac{\epsilon}{2N_2N_3}\} \, ,
\end{align}
where the sign in $(\bar{g}_{p,2}^{\star},\bar{g}_{p,3}^{\star},\bar{g}_{2}^{\star})$ is the upper one when $\bar{g}_S^{\star}=0$ and the lower one when $\bar{g}_S^{\star}=\epsilon/2$. 

The stability matrix is triangular at large $N_1$ and the critical exponents are given by the diagonal elements:
\begin{align}
\partial \beta_{S,D}(\bar{g}^{\star})&=\left\{\begin{array}{l}
-\epsilon \quad \text{ if } \bar{g}_{S,D}^{\star}=0 \\
 \epsilon \quad \text{ else}
\end{array}\right. \crcr
\partial \beta_{p,i}=&\left\{ \begin{array}{l}
 -\epsilon \quad \text{ if } (\bar{g}_S^{\star},\bar{g}_{p,i}^{\star})=(0,0) \text{ or } (\frac{\epsilon}{2},-\frac{\epsilon}{2N_j}) \text{ with } (i,j)=\{(2,3),(3,2)\} \\
 \epsilon \quad \text{ else}
\end{array} \right. \crcr
\partial \beta_{2}=&\left\{ \begin{array}{l}
 -\epsilon \quad \text{ if } (\bar{g}_S^{\star},\bar{g}_{2}^{\star})=(0,0) \text{ or } (\frac{\epsilon}{2},-\frac{\epsilon}{2N_2N_3})  \\
 \epsilon \quad \text{ else} 
\end{array} 
\right.
\end{align}

The only stable fixed point in all five directions is: 
$(\bar{g}_S^{\star},\bar{g}_D^{\star},\bar{g}_{p,2}^{\star},\bar{g}_{p,3}^{\star},\bar{g}_2^{\star})=(\frac{\epsilon}{2},-\frac{\epsilon}{2},0,0,0).$
This corresponds to $\bar{g}_{p,1}=\frac{\epsilon}{2}$ and $\bar{g}^{\star}=\bar{g}_{p,2}^{\star}=\bar{g}_{p,3}^{\star}=\bar{g}_2^{\star}=0$. It is a chiral fixed point with symmetry $O(N_1)\times O(N_2N_3)$, similar to those found in bi-fundamental models $O(N)\times O(M)$.

\medskip

In summary, we find \emph{no} real stable fixed point with non-zero tetrahedral coupling in the vector like limit.

\subsection{Matrix-like limit}
\label{sec:short-range_matrix}

We now consider the matrix-like double-scaling large-$N$ limit:
\begin{equation}
N_1=c N \,, \; N_2=  N \,, \; N \rightarrow \infty \,,
\end{equation}
with $N_3$ fixed and $c\ge 1$ fixed and of order one. We redefine the double-trace coupling, combining it with the third pillow coupling: 
\begin{align}
\tilde{g}_{dp}=\tilde{g}_d+\frac{\tilde{g}_{p,3}}{N_3} \,,
\end{align}
and we scale all the couplings with $N$ as:
\begin{equation}
\tilde{g}=\frac{\bar{g}}{N} \,, \; \tilde{g}_{p,1}=\frac{\bar{g}_{p,1}}{N} \,, \; \tilde{g}_{p,2}=\frac{\bar{g}_{p,2}}{N} \,, \; \tilde{g}_{p,3}=\frac{\bar{g}_{p,3}}{N^2} \,, \; \tilde{g}_{dp}=\frac{\bar{g}_{dp}}{N^2} \,. 
\end{equation}

The barred couplings are 't Hooft couplings, fixed in the large-$N$ limit, and we recognize the standard scaling of quartic matrix invariants with a single trace (the tetrahedron and first two pillows) or a double trace (the third pillow and the double-trace). The third pillow behaves effectively as a double-trace because the vertical line in Fig.~\ref{fig:interactions} corresponds in this case to the index whose range remains finite (i.e.\ $N_3$).

The one-loop beta functions at leading order in $1/N$ are:
\begin{align}
\beta_t&= -\epsilon \bar{g} +4\bar{g}\left(c\bar{g}_{p,1}+\bar{g}_{p,2}\right) \,,\crcr
\beta_{p,1}&= -\epsilon \bar{g}_{p,1}+2c\left(\bar{g}_{p,1}^2+\bar{g}^2\right)+4\bar{g}_{p,1}\left(\bar{g}+\bar{g}_{p,2}\right)+2N_3\bar{g}_{p,1}^2 \,, \crcr
\beta_{p,2}&= -\epsilon \bar{g}_{p,2}+2\left(\bar{g}_{p,2}^2+\bar{g}^2\right)+4c\bar{g}_{p,2}\left(\bar{g}+\bar{g}_{p,1}\right)+2cN_3\bar{g}_{p,2}^2 \,, \crcr
\beta_{p,3}&= -\epsilon \bar{g}_{p,3} +8\bar{g}_{p,1}\bar{g}_{p,2}+4
\bar{g}_{p,3}\left(c\bar{g}_{p,1}+\bar{g}_{p,2}+(1+c)\bar{g}\right)+2c\bar{g}_{p,3}^2+8\bar{g}\left(\bar{g}_{p,1}+\bar{g}_{p,2}\right)+2(N_3+2)\bar{g}^2 \,,\crcr
\beta_{dp}&=-\epsilon \bar{g}_{dp}+4\bar{g}_{dp}(1+c)\bar{g}+2cN_3\bar{g}_{dp}^2+\frac{2(N_3+2)}{N_3}\bar{g}^2+6\left(\bar{g}_{p,1}^2+\bar{g}_{p,2}^2\right)+\frac{4(2+N_3^2)}{N_3}\bar{g}_{p,1}\bar{g}_{p,2}\crcr
& +4\bar{g}_{dp}\left((c+N_3)\bar{g}_{p,1}+(1+N_3c)\bar{g}_{p,2}\right)+\frac{4(2+N_3)}{N_3}\bar{g}\left(\bar{g}_{p,1}+\bar{g}_{p,2}\right)\,.
\end{align}

We find 32 fixed points. For $\bar{g}^{\star},\bar{g}_{p,1}^{\star},\bar{g}_{p,2}^{\star}$, we find either:
\begin{align}
\bar{g}^{\star}&=0  \,,\crcr
(\bar{g}_{p,1}^{\star},\bar{g}_{p,2}^{\star})&= \{ (0,0),(0,\frac{\epsilon}{2(1+cN_3)}),(\frac{\epsilon}{2(c+N_3)},0) \,,\crcr
&(-\frac{\epsilon(1-cN_3)}{2\left(c^2N_3+c(N_3^2-3)+N_3\right)},-\frac{\epsilon(c-N_3)}{2\left(c^2N_3+c(N_3^2-3)+N_3\right)}) \} \,,
\label{eq:fpmatrix0}
\end{align}
or
\begin{align}
\bar{g}^{\star}&=\frac{\epsilon\left(c(1-N_3)\pm(1-c)N_3\sqrt{c^2+1-cN_3}\right)}{4(1+c)(c+N_3(c-1)^2)}  \,,\crcr
\bar{g}_{p,1}^{\star}&=\frac{\epsilon\left(cN_3(c-1)+1 \mp \sqrt{c^2+1-cN_3}\right)}{4(1+c)(c+N_3(c-1)^2)} \,,\crcr
\bar{g}_{p,2}^{\star}&=\frac{\epsilon\left(N_3(1-c)+c^2 \pm c\sqrt{c^2+1-cN_3}\right)}{4(1+c)(c+N_3(c-1)^2)} \,,
\label{eq:fpmatrix1}
\end{align}
or
\begin{align}
\bar{g}^{\star}&=\frac{\epsilon\left(c(1+c)(1-N_3)\pm (N_3(c^2+1)-2c)\sqrt{c^2+1-cN_3}\right)}{4N_3(c^2+1)^2-4c(3c^2-2c+3)}  \,,\crcr
\bar{g}_{p,1}^{\star}&=\frac{\epsilon\left(c^3N_3-2c^2+c(N_3+1)-1 \pm (c-1)\sqrt{c^2+1-cN_3}\right)}{4N_3(c^2+1)^2-4c(3c^2-2c+3)} \,, \crcr
\bar{g}_{p,2}^{\star}&=\frac{\epsilon\left(N_3-2c+c^2(N_3+1)-c^3 \mp c(c-1)\sqrt{c^2+1-cN_3}\right)}{4N_3(c^2+1)^2-4c(3c^2-2c+3)} \,,
\label{eq:fpmatrix2}
\end{align}
where the signs are taken to be simultaneously either the upper or the lower ones. 

For the last two couplings, we find the following fixed points in terms of $\bar{g}^{\star},\bar{g}_{p,1}^{\star},\bar{g}_{p,2}^{\star}$:
\begin{align}
\bar{g}_{p,3}^{\star}&=\frac{1}{4c}\Bigg[ \epsilon-4c\bar{g}_{p,1}^{\star}-4\bar{g}_{p,2}^{\star}-4(1+c)\bar{g}^{\star} \crcr
& \pm \sqrt{\left(4c(\bar{g}_{p,1}^{\star}+\bar{g}^{\star})+4(\bar{g}_{p,2}^{\star}+\bar{g}^{\star})-\epsilon\right)^2-16c\left(4(\bar{g}^{\star}+\bar{g}_{p,1}^{\star})(\bar{g}^{\star}+\bar{g}_{p,2}^{\star})+\bar{g}^{\star}{}^2(N_3-2)\right)}\Bigg] \crcr
\bar{g}_{dp}^{\star}&=\frac{1}{4cN_3}\Bigg[\epsilon-4c(\bar{g}_{p,1}^{\star}+\bar{g}^{\star}+N_3\bar{g}_{p,2}^{\star})-4(\bar{g}_{p,2}^{\star}+\bar{g}^{\star}+N_3\bar{g}_{p,1}^{\star})  \,,\crcr
& \pm \Big(16c^2(\bar{g}_{p,1}^{\star}+\bar{g}^{\star}+N_3\bar{g}_{p,2}^{\star})^2+(4(\bar{g}_{p,2}^{\star}+\bar{g}^{\star}+N_3\bar{g}_{p,1}^{\star})-\epsilon)^2-32c\left(\bar{g}_{p,1}^{\star}\bar{g}_{p,2}^{\star}+\bar{g}^{\star}(\bar{g}_{p,1}^{\star}+\bar{g}_{p,2}^{\star})\right)\crcr
& \qquad -16cN_3(\bar{g}_{p,1}^{\star}{}^2+\bar{g}_{p,2}^{\star}{}^2+\bar{g}^{\star}{}^2)-8\epsilon c(\bar{g}_{p,1}^{\star}+\bar{g}^{\star}+N_3\bar{g}_{p,2}^{\star})\Big)^{1/2}\Bigg] \,,
\label{eq:fpmatrixp3d}
\end{align}
where the signs are chosen independently. Notice that since $\bar{g}^{\star},\bar{g}_{p,1}^{\star},\bar{g}_{p,2}^{\star}$ are of order $\epsilon$, so are also $\bar{g}_{p,3}^{\star},\bar{g}_{dp}^{\star}$.

We can now compute the critical exponents. The stability matrix is a block triangular matrix with a first block corresponding to the couplings $\bar{g}^{\star},\bar{g}_{p,1}^{\star},\bar{g}_{p,2}^{\star}$ and a second diagonal block for the couplings $\bar{g}_{p,3}^{\star},\bar{g}_{dp}^{\star}$. 

We first compute the critical exponents for the couplings $\bar{g}^{\star},\bar{g}_{p,1}^{\star},\bar{g}_{p,2}^{\star}$. For the fixed points of equation \ref{eq:fpmatrix0}, we find the following critical exponents:

\begin{align}
(\omega_t,\omega_1,\omega_2)&=\{ (-\epsilon,-\epsilon,-\epsilon),(\epsilon,\frac{(1-cN_3)\epsilon}{1+cN_3},\frac{(1-cN_3)\epsilon}{1+cN_3}),(\epsilon,\frac{(c-N_3)\epsilon}{c+N_3},\frac{(c-N_3)\epsilon}{c+N_3}),\crcr
&(\epsilon,-\frac{(1-cN_3)(c-N_3)\epsilon}{c^2N_3+N_3+c(N_3^2-3)},\frac{(1-cN_3)(c-N_3)\epsilon}{c^2N_3+N_3+c(N_3^2-3)}) \} \,.
\end{align}

For the fixed points of equation \ref{eq:fpmatrix1} we find:
\begin{align}
\omega_t&=\epsilon\crcr
\omega_1&=-\omega_2=\frac{\epsilon\sqrt{c^2+1-cN_3}}{(1+c)(c+(c-1)^2N_3)}\Bigg(c^2(1-2N_3)-c(c-1)^2N_3^3+N_3^2(c^2-c+1)^2\crcr
&\pm 2N_3(N_3-1)c(c-1)\sqrt{c^2+1-cN_3}\Bigg)^{1/2}\,.
\end{align}

For the fixed points of equation \ref{eq:fpmatrix2}, we find:
\begin{align}
\omega_t&=\epsilon \crcr
\omega_1&=\omega_2=\epsilon\frac{c(N_3^2+2)(c^2+1)-N_3(c^4+4c^2+1) \pm c(N_3-1)(1+c)\sqrt{c^2+1-cN_3} }{(c^2+1)^2N_3-c(3c^2-2c+3)}\,.
\label{eq:critmatrix3}
\end{align}

We are interested in finding stable fixed points with non zero tetrahedron coupling.
The fixed points in \eqref{eq:fpmatrix0} have zero tetrahedral coupling and the ones in equation \eqref{eq:fpmatrix1} have $\omega_1=-\omega_2$ hence cannot be stable in all five directions.

We are left with the fixed points of equation \eqref{eq:fpmatrix2}. Because of the square root, for these fixed points $\omega_{1,2}$ are real only for $N_3\leq \frac{c^2+1}{c}$. For these values of $N_3$, the branch with a minus sign always has negative or zero critical exponents. The solutions with a plus sign is positive for $c\le N_3\leq \frac{c^2+1}{c}$. If $N_3 > \frac{c^2+1}{c}$, both solutions are complex with a positive real part.

We thus have to look at the critical exponents for the last two couplings in order to conclude. As the second block of the stability matrix is diagonal, the critical exponents for the last two couplings are just the diagonal elements:

\be
\partial \beta_{p,3}(\bar{g}^{\star})=\pm \frac{\epsilon  \sqrt{R_1\pm R_2\sqrt{c^2+1-cN_3}}}{|(c^2+1)^2N_3-c(3c^2-2c+3)|} \,, \qquad
\partial \beta_{dp}(\bar{g}^{\star})=\pm \frac{\epsilon \sqrt{R_3 \pm R_4\sqrt{c^2+1-cN_3}}}{|(c^2+1)^2N_3-c(3c^2-2c+3)|} \,,
\ee
with:
\begin{align}
R_1=& c^2(c^2+1)^2N_3^4-c\left(2c^6+11c^4+2c^3+11c^2+2\right)N_3^3\crcr
& +\left(c^8+9c^6+20c^5+24c^4+20c^3+9c^2+1\right)N_3^2\crcr
& +c\left(4c^6-22c^5-5c^4-54c^3-5c^2-22c+4\right)N_3\crcr
& -c^2\left(3c^4-30c^3+14c^2-30c+3\right) \,, \crcr
R_2=& 2c\left(c+1\right)\Big(c(c^2+1)N_3^3-\left(3c^4+c^3+8c^2+c+3\right)N_3^2\crcr
& \qquad +\left(5c^4+16c^2+5\right)N_3-6c(c^2+1)\Big) \,, \crcr
R_3=&3c^2(c^2+1)^2N_3^4-3c\left(2c^6+11c^4+2c^3+11c^2+2\right)N_3^3\crcr
& +\left(c^8+31c^6+12c^5+84c^4+12c^3+31c^2+1\right)N_3^2\crcr
& -c^2\left(14c^4+39c^3+34c^2+39c+14\right)N_3\crcr
& -c^2\left(3c^4-30c^3+14c^2-30c+3\right) \,,\crcr
R_4=& 6c^2\left(c^2+1\right)\left(c+1\right)\left(N_3-1\right)\left(N_3-\frac{2c}{c^2+1}\right)\left(N_3-\frac{c^2+1}{c}\right) \,.
\end{align}
and the signs in front of  $\partial \beta_{p,3}(\bar{g}^{\star})$ and $\partial \beta_{dp}(\bar{g}^{\star})$ are the same as in $\bar{g}_{p,3}^{\star}$ and $\bar{g}_{dp}^{\star}$ while the signs inside the square roots are taken to be simultaneously the same as in \ref{eq:fpmatrix2}.

For $c \le N_3\leq \frac{c^2+1}{c}$, $\partial \beta_{p,3}(\bar{g}^{\star})$ can be real and positive, but in this case $\partial \beta_{dp}(\bar{g}^{\star})$ is purely imaginary.

However, for $N_3 > \frac{c^2+1}{c}$, $\partial \beta_{p,3}(\bar{g}^{\star})$ and $\partial \beta_{dp}(\bar{g}^{\star})$ are both complex and we can choose the sign in front so that the real part is positive. 

\medskip

Summarizing the findings of this subsection: we find \emph{no} real stable fixed point in the matrix-like large-$N$ limit; however, for $N_3>\frac{c^2+1}{c}$ we do find a complex infrared fixed point  stable in all the five directions.

\subsection{Tensor-like limit}
\label{sec:short-range_triple}
We finally consider the large-$N$  limit with:
	\begin{align}
		N_1 \, = \, N_2  = \, N_3  =  \, N \, , \quad {\rm and} \quad N \, \to \, \infty \, .
	\end{align}
While we could, like in the previous sections, consider an inhomogeneous scaling with ratios different from one (e.g.\ $N_1/N_2=c$) this leads to very bulky formulas, with no much qualitative gain. We will thus stick to the homogeneous case. 

The resulting $O(N)^3$ model has been studied at leading order in $1/N$ \cite{Giombi:2017dtl}. Here we analyze the fate of its fixed points at subleading orders in $1/N$. We combine the three pillow couplings into one coupling $\tilde{g}_p/3=\tilde{g}_{p,1}=\tilde{g}_{p,2}=\tilde{g}_{p,1}$, thus endowing the model with a discrete color permutation symmetry. 
After scaling the couplings as:
	\begin{align} \label{eq:bar_g}
		\gt \, = \, \frac{\bar{g}}{N^{3/2}} \, , \qquad \gt_p \, = \, \frac{\bar{g}_p}{N^2} \, , \qquad \gt_d \, = \, \frac{\bar{g}_d}{N^3} \, , 
	\end{align}
the two-loop beta functions up to order $\cO(N^{-3/2})$ are:
	\begin{align} \label{eq:beta_t-SR}
		\bar{\beta}_t \, &= \, - \, \epsilon\bar{g} \, + \, 2\bar{g}^3 \, +\frac{2\bar{g}\bar{g}_p}{3N}\Big[6-\bar{g}_p\Big] \, + \, \mathcal{O}(N^{-3/2}) \, , \\
		\bar{\beta}_{p} \, &= \, - \, \epsilon\bar{g}_{p} \, + \, 6\bar{g}^2 +  \, \frac{2\bar{g}_{p}^2}{3}
		\, - \, 2\bar{g}^2 \bar{g}_{p}   + \, \frac{8\bar{g}}{N^{1/2}} \Big[  \bar{g}_{p} - 3\bar{g}^2  \Big] \nonumber \\
		& \quad + \frac{2}{9N} \Big[5\bar{g}_p^2\left(3-\bar{g}_p\right)+54\bar{g}^2\left(1-2\bar{g}_p\right)\Big]\, + \, \mathcal{O}(N^{-3/2}) \, , \nonumber\\
		\bar{\beta}_d \, &= \, - \, \epsilon\bar{g}_d \, + \,\frac{2}{3} \left(  3\bar{g}_d^2 + 6 \, \bar{g}_{p} \bar{g}_d + 2 \bar{g}_{p}^2 \right)  - 2\bar{g}^2 \left( 5\bar{g}_d + 4 \bar{g}_{p} \right) \\ 
	& \quad \, + \frac{4\bar{g}}{N^{1/2}} \Big[ \bar{g}_{p} +  3\bar{g}_d -3\bar{g}^2  \Big] \,  \nonumber \\
	& \quad +\frac{2\bar{g}_p}{9N}\Big[18\bar{g}_d+9\bar{g}_p-15\bar{g}_d\bar{g}_p-14\bar{g}_p^2-36\bar{g}^2 \Big] \, + \, \mathcal{O}(N^{-3/2}) \, . \nonumber
	\end{align}
At leading order we reproduce the results obtained in \cite{Giombi:2017dtl}.

We remark that, while at leading order the tetrahedral beta function has no quadratic term, such a term appears at the first non-zero subleading order, $N^{-1}$. In order to better understand the implications of this, consider the fictitious single-coupling beta function $-\epsilon g + g^3 +\frac{2 a}{N} g^2$, with $a$ some real constant; its fixed points are:
\begin{equation} \label{eq:ex-FP}
g_{\star,\pm} = -\frac{a}{N} \pm \sqrt{\epsilon+\frac{a^2}{N^2}} \,.
\end{equation}
If we expand at large $N$, we find:
\be
g_{\star,\pm} = \pm \sqrt{\epsilon} -\frac{a}{N} \pm \frac{a^2}{2\sqrt{\epsilon} N^2} + \cO(N^{-3})\,,
\ee
and naively we seem to have a problem: the subleading orders are non-perturbative and even blow up for $\epsilon\to 0$. We would thus conclude that the $\sqrt{\epsilon}$ fixed point is spurious.
However, a more careful look reveals that the behavior of the fixed points \eqref{eq:ex-FP} is actually governed by the combination $\epsilon N^2$.
For $\epsilon N^2 \ll 1$ (towards the finite $N$ range), the usual one-loop-driven Wilson-Fisher fixed point is obtained, $g_{\star,+}\sim N\epsilon/2a$. For $\epsilon N^2 \gg 1$, one gets instead the two-loop-driven fixed points typical of the $O(N)^3$ model in the melonic limit \cite{Giombi:2017dtl}, $\bar{g}_{\star,\pm}\sim \pm \sqrt{\epsilon}$. As we wish to study the $1/N$ corrections to the melonic leading order, we need to assume $\epsilon N^2 \gg 1$. To this end, we set:
\be
N=\tilde{N}/\sqrt{\epsilon}\,,
\ee
and we expand beta functions, fixed points and critical exponents in $1/\tilde{N}$ first, and only afterwards in $\epsilon$. 
The beta functions become:
	\begin{align}
		\bar{\beta}_t \, &= \, - \, \epsilon\bar{g} \, + \, 2\bar{g}^3 \, +\frac{2\sqrt{\epsilon}\bar{g}\bar{g}_p}{3\tilde{N}}\Big[6-\bar{g}_p\Big] \, + \, \mathcal{O}(\tilde{N}^{-3/2}) \, , \\
		\bar{\beta}_{p} \, &= \, - \, \epsilon\bar{g}_{p} \, + \, 6\bar{g}^2 +  \, \frac{2\bar{g}_{p}^2}{3}
		\, - \, 2\bar{g}^2 \bar{g}_{p}   + \, \frac{8\epsilon^{1/4}\bar{g}}{\tilde{N}^{1/2}} \Big[  \bar{g}_{p} - 3\bar{g}^2  \Big] \nonumber \\
		& \quad + \frac{2\sqrt{\epsilon}}{9\tilde{N}} \Big[5\bar{g}_p^2\left(3-\bar{g}_p\right)+54\bar{g}^2\left(1-2\bar{g}_p\right)\Big]\, + \, \mathcal{O}(\tilde{N}^{-3/2}) \, , \nonumber\\
		\bar{\beta}_d \, &= \, - \, \epsilon\bar{g}_d \, + \,\frac{2}{3} \left(  3\bar{g}_d^2 + 6 \, \bar{g}_{p} \bar{g}_d + 2 \bar{g}_{p}^2 \right)  - 2\bar{g}^2 \left( 5\bar{g}_d + 4 \bar{g}_{p} \right) \\ 
	& \quad \, + \frac{4\epsilon^{1/4}\bar{g}}{\tilde{N}^{1/2}} \Big[ \bar{g}_{p} +  3\bar{g}_d -3\bar{g}^2  \Big] \,  \nonumber \\
	& \quad +\frac{2\sqrt{\epsilon}\bar{g}_p}{9\tilde{N}}\Big[18\bar{g}_d+9\bar{g}_p-15\bar{g}_d\bar{g}_p-14\bar{g}_p^2+36\bar{g}^2 \Big] \, + \, \mathcal{O}(\tilde{N}^{-3/2}) \, . \nonumber
	\end{align}
	
We parametrize the critical couplings as:
\begin{align}
		\bar{g}^\star \,& = \, \bar{g}^\star_{(0)} \, + \, \tilde{N}^{-\frac{1}{2}} \, \bar{g}^\star_{(1)} \, +\tilde{N}^{-1}\bar{g}^{\star}_{(2)}\, + \, \mathcal{O}(\tilde{N}^{-3/2}) \, , \crcr
\bar{g}^\star_p \, &= \, \bar{g}^\star_{p,(0)} \, + \, \tilde{N}^{-\frac{1}{2}} \, \bar{g}^\star_{p,(1)} \, +\tilde{N}^{-1}\bar{g}^{\star}_{p,(2)}\, + \, \mathcal{O}(\tilde{N}^{-3/2}) \, , \crcr
\bar{g}^\star_d \, &= \, \bar{g}^\star_{d,(0)} \, + \, \tilde{N}^{-\frac{1}{2}} \, \bar{g}^\star_{d,(1)} \, +\tilde{N}^{-1}\bar{g}^{\star}_{d,(2)}\, + \, \mathcal{O}(\tilde{N}^{-3/2}) \, .
\label{eq:param_couplings}
	\end{align}
Solving for the zeros of the beta functions at leading order we find the following complex solutions\footnote{There are also an additional four solutions with zero tetrahedral coupling:
\[
(\bar{g}_{p,(0)}^{\star},\bar{g}_{d,(0)})=\{(0,0),(0,\frac{\epsilon}{2}),(\frac{3\epsilon}{2},-\frac{3\epsilon}{2}),(\frac{3\epsilon}{2},-\epsilon)\} \,.
\]
We do not study them further as we are interested in fixed points with non-zero tetrahedral coupling.}:
	\begin{align} \label{eq:LO-FP-SR}
		\bar{g}^\star_{(0)} \, = \, \pm \, \sqrt{ \frac{\epsilon}{2}} \, , \qquad 
		\bar{g}_{p,(0)}^\star \, = \, \pm \, 3 i \sqrt{ \frac{\epsilon}{2}} \, +\frac{3\epsilon}{2} +\mathcal{O}(\epsilon^{3/2})\, , \qquad 
		\bar{g}_{d,(0)}^\star \, = \, \mp \,  i \sqrt{ \frac{\epsilon}{2}}(3\pm \sqrt{3}) +\mathcal{O}(\epsilon^{3/2})\, .
	\end{align}
There are eight solutions by the combination of the signs in $\bar{g}^\star_{(0)}$, $\bar{g}_{p,(0)}^\star$ and the relative sign of $3$ and $\sqrt{3}$ in $\bar{g}_{d,(0)}^\star$.
The overall sign of $\bar{g}_{d,(0)}^\star$ is associated with the sign of $\bar{g}_{p,(0)}^\star$.

Next we compute the subleading $\mathcal{O}(\tilde{N}^{-1/2})$ corrections to the fixed points by substituting \eqref{eq:param_couplings}, with the leading order given by \eqref{eq:LO-FP-SR}, into the beta functions and solving for the $\mathcal{O}(\tilde{N}^{-1/2})$ order. Since $\bar{\beta}_t$ does not have a $\mathcal{O}(\tilde{N}^{-1/2})$ contribution, the equation for $\bar{g}^\star_{(1)}$ comes only from the leading order part, evaluated at linear order in the coupling correction, leading to:
	\begin{align}
		\bar{g}^\star_{(1)} \, = \, 0 \, , \qquad ({\rm for} \ \epsilon \, \ne \, 0) \,.
	\end{align}
For the pillow and double-trace we find instead non-trivial corrections: 
	\begin{align}
		\bar{g}^\star_{p,(1)} \, = \, \mp \, 3 \sqrt{2}\epsilon^{3/4} \, , \qquad 
		\bar{g}^\star_{d,(1)} \, = \, \pm \, 3 \frac{\epsilon^{3/4}}{\sqrt{2}} \, .
	\end{align}
The choice of upper or lower sign for $\bar{g}^\star_{p,(1)}$ and $\bar{g}^\star_{d,(1)}$ is synchronized with that for $\bar{g}_{(0)}^\star$.
At next order, $\tilde{N}^{-1}$, we find:\footnote{We checked the next two orders, and we have found that the order $\tilde{N}^{-3/2}$ starts again at order $\epsilon^{3/4}$, and order $\tilde{N}^{-2}$ at order $\sqrt{\epsilon}$. We do not know whether this pattern repeats to all orders.}
\begin{align}
		\bar{g}^{\star}_{(2)} \, &= \, \pm 3i\sqrt{\epsilon}  \mp \frac{9\epsilon}{2\sqrt{2}}+ \mathcal{O}(\epsilon^{3/2})\, , \crcr
		\bar{g}^\star_{p,(2)} \, &= \,  9\sqrt{\epsilon} \mp \frac{33i \epsilon}{\sqrt{2}}+ \mathcal{O}(\epsilon^{3/2}) \, , \crcr
		\bar{g}^\star_{d,(2)} \, &= \, - \, 3 \sqrt{3\epsilon}\left(\sqrt{3}\pm 1\right) \pm \frac{3\sqrt{2}i\epsilon}{2}\left(5\pm 2\sqrt{3}\right)+ \mathcal{O}(\epsilon^{3/2}) \, .
\end{align}
where the first sign in $\bar{g}^{\star}_{(2)}$ is the upper one when the choice of sign was the same for $\bar{g}^{\star}_{(0)}$ and $\bar{g}^{\star}_{p,(0)}$, the second sign in $\bar{g}^{\star}_{(2)}$ is synchronized with the sign of $\bar{g}^{\star}_{(0)}$, the signs in $\bar{g}^{\star}_{p,(2)}$ and in front of the $\epsilon$ term in $\bar{g}^{\star}_{d,(2)}$ are synchronized with the sign of $\bar{g}^{\star}_{p,(0)}$ and the signs in the parenthesis in $\bar{g}^\star_{d,(2)}$ are synchronized with the sign in $\bar{g}^\star_{d,(0)}$.

We then compute the critical exponents up to order $\tilde{N}^{-1}$ to find:
\begin{align}
\omega_t&=2\epsilon \mp \frac{6 i \sqrt{2}\epsilon}{\tilde{N}}+\mathcal{O}(\epsilon^{3/2},\tilde{N}^{-3/2})  \,,\crcr
\omega_p&=\pm 2 i \sqrt{2\epsilon} + 12\frac{\sqrt{\epsilon}\mp i \sqrt{2}\epsilon}{\tilde{N}} +\mathcal{O}(\epsilon^{3/2},\tilde{N}^{-3/2}) \,, \crcr
\omega_d&= \pm 2i \sqrt{6\epsilon} \mp 12\sqrt{3} \frac{\sqrt{\epsilon}\mp i \sqrt{2}\epsilon}{\tilde{N}} +\mathcal{O}(\epsilon^{3/2},\tilde{N}^{-3/2})  \,.
\end{align}

The sign of the leading order in $\omega_p$ is the same one as in $\bar{g}_{p,(0)}^{\star}$. The sign of the leading order in $\omega_d$ is the upper one if the choices of signs in $\bar{g}_{p,(0)}^{\star}$ and $\bar{g}_{d,(0)}^{\star}$ are different and the lower one if they are the same. The sign in front of the $\tilde{N}^{-1}$ term of $\omega_d$ is synchronized with $\bar{g}_{d,(0)}^{\star}$. The other signs for the orders $\tilde N^{-1}$ are synchronized with $\bar{g}_{p,(0)}^{\star}$. In particular, for the fixed points with the lower choice of sign in $\bar{g}_{d,(0)}^{\star}$, the real parts of all three critical exponents are positive. 

\medskip

In conclusion, the complex fixed point of the short range $O(N)^3$ model found in \cite{Giombi:2017dtl} persists at subleading orders in $1/N$. Importantly, the order $\tilde{N}^{-1/2}$ corrections to the critical exponents are zero, but the order $\tilde{N}^{-1}$ endow them with a real part, meaning that the fixed point is infrared stable.

\section{The long-range tri-fundamental model}
\label{sec:long-range}

\subsection{The long-range multi-scalar model}
\label{sec:long-range_ms}
The long-range multi-scalar model with quartic interactions in dimension $d$ is defined by the action:
\begin{align}
		S[\phi]  \, &= \,  \int d^dx \, \bigg[ \frac{1}{2} \phi_\mba(x) ( - \partial^2)^{\zeta}\phi_{\mba}(x) +\frac{1}{2}\, \kappa_{\mba \mbb}\phi_{\mba}(x) \phi_{\mbb}(x) + 
		\frac{1}{4!} \, \lambda_{\mba \mbb \mbc \mbd}
		\phi_{\mba}(x) \phi_{\mbb}(x) \phi_{\mbc}(x) \phi_{\mbd}(x) \bigg] \, ,
	\end{align}
where the coupling $\lambda_{\mba\mbb\mbc\mbd}$ and the mass parameter $\kappa_{\mba\mbb}$ are symmetric tensors.
The indices take values from 1 to $\cN$.
The model is ``long range'' due to the non trivial power of the Laplacian 
$0< \zeta < 1$. We will use the renormalization scheme and notations of \cite{Benedetti:2020rrq}.

We treat the mass parameter $\kappa$ as a perturbation, hence the covariance (propagator) of the free theory is 
$ C_{\mba \mbb}(x,y) =  \delta_{\mba \mbb} \;  C(x-y) $, with:
\be\label{eq:cov} 
\begin{split}
 & C(x-y) = \int\frac{d^dp}{(2\pi)^d}\; e^{ - \im p (x-y) } C(p) = \f{\G\left(\f{d-2\z}{2}\right)}{2^{2\z}\pi^{d/2}\G(\z)} \; \f{1}{|x-y|^{d-2\z}}\,, \\
 & C(p) = \frac{1}{p^{2\zeta}}  = \frac{1}{\Gamma(\zeta)} \int_0^{\infty} d\alpha \;\alpha^{\zeta -1 } e^{- \alpha p^2} \,.
\end{split}
\ee 
The canonical dimension of the field is:
\be
\Delta_{\phi} = \frac{d-2\zeta}{2} \,,
\ee
therefore, the quartic interaction is irrelevant for $\z<d/4$ leading to mean-field behavior (as rigorously proved in \cite{Aizenman:1988}), while for $\z>d/4$ it is relevant and a non-trivial IR behavior is expected. The marginal case is $\z=d/4$. We will be interested in the weakly relevant case:
\be \label{eq:zeta-eps}
 \zeta = \frac{d+\epsilon}{4} \,,
\ee
with small $\epsilon$. The ultraviolet dimension of the field is thus fixed to $\Delta_{\phi}=\frac{d-\epsilon}{4}$. 

In order to renormalize the UV divergences we  use the zero momentum BPHZ subtraction scheme.
However, since we are working with a massless propagator, an infrared regulator is required. We introduce that by modifying the propagator as: 
\be\label{eq:param}
 C_{\mu}(p) = \frac{1}{(p^2 + \mu^2)^{\zeta}} = \frac{1}{\Gamma(\zeta)} \int_0^{\infty} d\alpha \;\alpha^{\zeta -1 } e^{- \alpha p^2 -\alpha \mu^2} \,,
\ee
for some mass parameter $\mu>0$.
Using the results of \cite{Benedetti:2020rrq}, we have up to two-loop order: 
\begin{align}
\beta_{\mba \mbb \mbc \mbd} &= -\epsilon \gt_{\mba \mbb \mbc \mbd} + \alpha_{D}\left(\gt_{\mba \mbb \mbe \mbf}\gt_{\mbe \mbf \mbc \mbd} + 2 \textrm{ terms} \right)  + \alpha_{S}\left(\gt_{\mba \mbb \mbe \mbf}\gt_{\mbe \mbg \mbh \mbc}\gt_{\mbf \mbg \mbh \mbd}+ 5 \textrm{ terms}\right) 
\label{eq:beta_abcd_alpha}\,,
\end{align}
\begin{align}
\beta^{(2)}_{\mbc\mbd}
&= - (d-2\Delta_{\phi} ) \rt_{\mbc \mbd} +\alpha_{D}  \big( \rt_{ \mbe \mbf}\gt_{\mbe \mbf \mbc \mbd} \big) + \alpha_{S} \big(\rt_{\mbe \mbf} \gt_{\mbe \mbg \mbh \mbc}\gt_{\mbf \mbg \mbh \mbd} \big) \,. 
\label{eq:beta2_abcd_alpha}
\end{align}
The running couplings have been rescaled as
$g_{\mba \mbb \mbc \mbd}= \left(4\pi\right)^{d/2}\Gamma(d/2) \, \tilde{g}_{\mba \mbb \mbc \mbd} $ and 
$r_{\mba \mbb }= \left(4\pi\right)^{d/2}\Gamma(d/2) \, 
\rt _{\mba \mbb} $,
and the alpha's are defined by
	\begin{align}
& \alpha_{D} \, = \, 1 +\frac{\epsilon}{2}\big[\psi(1)-\psi(\tfrac{d}{2}) \big]+\frac{\epsilon^2}{8}\left[\left(\psi(1)-\psi(\tfrac{d}{2})\right)^2+ \psi_1(1)-\psi_1(\tfrac{d}{2})\right] \,, \crcr
&\alpha_{S} \, = \,  2\psi( \tfrac{d}{4} ) - \psi( \tfrac{d}{2})-\psi(1)   +\frac{\epsilon}{4}\Big[\left[2\psi(\tfrac{d}{4})-\psi(\tfrac{d}{2})-\psi(1)\right]
\left[3\psi(1)-5\psi(\tfrac{d}{2})+2\psi(\tfrac{d}{4})\right]   \crcr
& \qquad \; + 3\psi_1(1) + 4\psi_1(\tfrac{d}{4})-7\psi_1(\tfrac{d}{2})  -4 J_0(\tfrac{d}{4}) \Big] \, \,,
 \label{eq:alphas}
\end{align}
with $\psi_i$ the polygamma functions of order $i$ and $J_0$ the sum
\begin{equation}
J_0(\tfrac{d}{4})=\frac{1}{\Gamma(\tfrac{d}{4})^2}\sum_{n \geq 1}\frac{\Gamma(n+\tfrac{d}{2})\Gamma(n+ \tfrac{d}{4})^2}{n(n!)\Gamma(\tfrac{d}{2}+2n)}\Big[2\psi(n+1)-\psi(n)-2\psi(n+\tfrac{d}{4})-\psi(n+\tfrac{d}{2})+2\psi(\tfrac{d}{2}+2n)\Big] \,.
\end{equation}

\subsection{Large-$N$ expansion of the long-range $O(N)^3$ tensor model}
\label{sec:long-range_equalN}
We now set $N_1=N_2=N_3=N$ and study the fixed points of the long-range $O(N)^3$ model at next-to-leading order in $1/N$. We use:
	\begin{align}
		\gt_{\mba \mbb \mbc \mbd} \, &= \, \gt \left(\delta^t_{\mba \mbb \mbc \mbd} + 5 \textrm{ terms} \right)
		\, + \,  \gt_{p} \left(\delta^{p}_{\mba \mbb; \mbc \mbd} + 5 \textrm{ terms} \right)
		\, + \, 2\gt_d \left(\delta^d_{\mba \mbb \mbc \mbd} + 2 \textrm{ terms} \right) \, ,
	\label{eq:couplingtot}
	\end{align}
where like before each $\mba$ is a triplet of indices $\mba=(a_1,a_2,a_3)$.
$\delta^t_{\mba \mbb \mbc \mbd}$ and $\delta^d_{\mba \mbb \mbc \mbd}$ are defined as in \ref{eq:deltas-nohat}, and 
\begin{equation}
\delta^p_{\mba \mbb ;\mbc \mbd}=\frac{1}{3}\sum_{i=1}^3 \delta^{p,i}_{\mba \mbb ; \mbc \mbd} \, .
\end{equation}

The beta functions up to two-loops are then:
\begin{align}
\beta_t=&-\epsilon \tilde{g}+\frac{4\alpha_D}{3}\Big[2\tilde{g}_p^2+18\tilde{g}\tilde{g}_d+3(N+1)\tilde{g}\tilde{g}_p\Big]\crcr
& +\frac{4\alpha_S}{9}\Big[27(3N+2)\tilde{g}^3 + 54 \left(N^3+14\right) \tilde{g}_d^2 \tilde{g} +3\left(N^3+9 N^2+51 N+53\right) \tilde{g}_p^2 \tilde{g} + 2\left(2 N^2+13 N+24\right)\tilde{g}_p^3 \crcr
& \qquad \qquad  +18\left(2 N^2+5 N+14\right) \tilde{g}_p \tilde{g}^2 +36\tilde{g}_d\left(4\tilde{g}_p^2+9N\tilde{g}^2+3(N^2+3N+3)\tilde{g}\tilde{g}_p\right)\Big] \, , \crcr
\beta_p=& -\epsilon \tilde{g}_p + \frac{2\alpha_D}{3}\Big[36\tilde{g}_p\tilde{g}_d+3(N+2)\left(3\tilde{g}+4\tilde{g}_p \right)\tilde{g}+(N^2+5N+12)\tilde{g}_p^2\Big] \crcr
& +\frac{4\alpha_S}{9}\Big[54(N^2+N+4)\tilde{g}^3 +54 \left(N^3+14\right) \tilde{g}_d^2 \tilde{g}_p +18\left(5 N^2+19 N+30\right) \tilde{g}_p^2 \tilde{g} \crcr
& \qquad \qquad  + \left(4 N^3+27 N^2+135 N+179\right)\tilde{g}_p^3 +9  \left(N^3+6 N^2+51 N+50\right)\tilde{g}_p \tilde{g}^2 \crcr
& \qquad \qquad +36 \tilde{g}_d \left( \left(4 N^2+8 N+15\right)\tilde{g}_p^2+3 (7 N+8) \tilde{g}_p \tilde{g} +9(N+2) \tilde{g}^2 \right)  \Big] \, , \crcr
\beta_d=& -\epsilon \tilde{g}_d+\frac{2\alpha_D}{3}\Big[ 3\left(N^3+8\right) \tilde{g}_d^2 +6\left(N^2+N+1\right) \tilde{g}_d \tilde{g}_p +18N \tilde{g}_d \tilde{g} + (2N+3)\tilde{g}_p^2+ 6 \tilde{g}\tilde{g}_p \Big] \crcr
& +\frac{4\alpha_S}{9}\Big[ 27N \tilde{g}^3+ 216 \tilde{g}_d^2 \left( \left(N^2+N+1\right)\tilde{g}_p+3N \tilde{g} \right)+18  \left(5 N^3+22\right)\tilde{g}_d^3 \crcr
& \qquad \qquad +9 \tilde{g}_d \left(\left(N^3+3 N^2+17 N+17\right)\tilde{g}_p^2 +12  \left(N^2+N+3\right)\tilde{g}_p \tilde{g}+3  \left(N^3+3 N+2\right)\tilde{g}^2\right)\crcr
& \qquad \qquad +72(N+1) \tilde{g}_p^2 \tilde{g}+7  \left(N^2+3 N+5\right)\tilde{g}_p^3+18 \left(N^2+N+4\right) \tilde{g}_p \tilde{g}^2\Big] \, , \crcr
\beta^{(2)}=& -(d-2\Delta_{\phi})\tilde{r}+2\alpha_D\tilde{r}\Big[3N\tilde{g}+(N^2+N+1)\tilde{g}_p+(N^3+2)\tilde{g}_d\Big] \crcr
& +2\alpha_S\tilde{r}\Big[36N\tilde{g}\tilde{g}_d+12(N^2+N+1)\tilde{g}_p\left(\tilde{g}_d+\tilde{g}\right)+6(N^3+2)\tilde{g}_d^2+3(N^3+3N+2)\tilde{g}^2\crcr
& \qquad \qquad +(N^3+3N^2+9N+5)\tilde{g}_p^2\Big] \, . 
\end{align}

\subsubsection{Fixed points}

We rescale the couplings as:
\begin{equation}
\tilde{g}=\frac{\bar{g}}{N^{3/2}} \,, \; \tilde{g_{p}}=\frac{\bar{g_{p}}}{N^2} \,, \; \tilde{g_d}=\frac{\bar{g}}{N^3} \,,
\end{equation}
and first consider the large $N$ limit.

In \cite{Benedetti:2019eyl} it was found that at $\epsilon=0$ the tetrahedron coupling $\bar{g}$ is exactly marginal in the large-$N$ limit, 
and it parametrizes a line of fixed points for the remaining two couplings. The exact marginality is due to the fact that at large $N$ the tetrahedron receives no radiative corrections, and moreover in the long-range case there is no wave-function renormalization. The latter is responsible for the $2\bar{g}^3$ term in \eqref{eq:beta_t-SR}, which is absent in the long-range case.
However, at order $N^{-1}$ the tetrahedron beta function is non-zero also in the long-range model, and excluding uncontrolled non-perturbative fixed points, the line of fixed points collapses to the trivial fixed point at vanishing couplings.
Turning on $\epsilon$ does not help, as it contributes a term $-\epsilon \bar{g}$, that being the only term of order $N^0$, leads to $\bar{g}^\star=0$ already at leading order. As we did before, it is instructive to consider again a fictitious single-coupling beta function to guide our understanding; the situation we have in the long range model, at $\epsilon\neq 0$ and at next-to-leading order in $1/N$, is captured by a beta function of the form $-\epsilon g + g^2/N$. Its fixed points are the trivial one, and $g^\star= N\epsilon$, which goes to infinity if we take $N\to\infty$ at fixed $\epsilon$.
Similarly to what we have seen in the short-range case, the problem is resolved by specifying how small should $\epsilon$ be in comparison to  $1/N$. In particular, it is clear that we now need $ N \epsilon \ll 1$. In other words, we should move the $-\epsilon \bar{g}$ term to the first non-trivial order in $1/N$, by setting
\be
\epsilon=\frac{\tilde{\epsilon}}{N}\,,
\ee
and expanding as before in $1/N$ first, and then in $\tilde{\epsilon}$.
Notice that the condition $ N \epsilon \ll 1$ is compatible with the $N \sqrt{\epsilon}\gg 1$ condition which we had in the short-range case. Of course the meaning of $\epsilon$ is different in the two cases, but in practice their role is similar.
We also note that a similar tuning of $\epsilon$ and $N$ was considered in \cite{Fleming:2020qqx} in order to find a finite-$N$ precursor of the large-$N$ line of fixed points in the $O(N)$ model with $(\phi^2)^3$ interaction.

To simplify the computations we define two new independent couplings as in \cite{Benedetti:2019eyl}:

\begin{equation} \label{eq:g_1,2}
\bar{g}_1=\frac{\bar{g}_p}{3} \,, \; \bar{g}_2=\bar{g}_d+\bar{g}_p\,.
\end{equation}

Parametrizing the coefficients of the $\epsilon$ expansion  of the one- and two-loop constants $\alpha$ as: 
	\begin{align}
		\alpha_D \, &= \, 1\, + \, \alpha_{D,1} \, \epsilon \, + \, \alpha_{D,2} \, \epsilon^2 \, + \, \mathcal{O}(\epsilon^3) \, ,\crcr
		\alpha_S \, &= \, \alpha_{S,0} \, + \, \alpha_{S,1} \, \epsilon \,  + \, \mathcal{O}(\epsilon^2) \, ,
	\label{eq:alpha param}
	\end{align}
the beta functions at two loops up to order $N^{-1}$ are:

\begin{align}
\beta_t \, = & \, \frac{\bar{g}}{N}\Big[12\bar{g_1}\left(1+\alpha_{S,0} \bar{g_1} \right) -\tilde{\epsilon}\Big] +\mathcal{O}(N^{-3/2})\,, \crcr
\beta_1=& 2\left(\bar{g}_1^2+\bar{g}^2\right)+4\alpha_{S,0}\bar{g}_1\bar{g}^2 +\frac{8\bar{g}}{N^{1/2}} \Big[ \bar{g}_1 + \alpha_{S,0} \bar{g}^2 \Big]  \crcr 
&+\frac{1}{N}\bigg[  10\bar{g}_1^2+4\bar{g}^2+8\alpha_{S,0}\bar{g}_1\left(2\bar{g}_1^2+3\bar{g}^2\right) +\tilde{\epsilon}\left(2\alpha_{D,1}(\bar{g}_1^2+\bar{g}^2)+\bar{g}_1(4\alpha_{S,1}\bar{g}^2-1)\right)\bigg] +\mathcal{O}(N^{-3/2})\,, \crcr
\beta_2=& 2\left(\bar{g_2}^2+3\bar{g}^2\right)+12\alpha_{S,0}\bar{g_2}\bar{g}^2 +\frac{12\bar{g}}{N^{1/2}} \Big[ \bar{g_2} + 3\alpha_{S,0} \bar{g}^2 \Big] \crcr 
&+\frac{1}{N} \bigg[ 12\left(\bar{g_1}^2+\bar{g}^2+\bar{g_1}\bar{g_2}\right)+12\alpha_{S,0}\bar{g_1}\left(2\bar{g_1}^2+3\bar{g_2}\bar{g_1}+8\bar{g}^2\right)\crcr
& +\tilde{\epsilon}\left(2\alpha_{D,1}(\bar{g}_2^2+3\bar{g}^2)+\bar{g}_2(12\alpha_{S,1}\bar{g}^2-1)\right)
 \bigg] +\mathcal{O}(N^{-3/2})\,, \crcr
\beta^{(2)}=&-\frac{d}{2}\tilde{r}+2\left(2\bar{g}_2+3\alpha_{S,0}\bar{g}^2\right)\tilde{r} +\frac{6\bar{g}\tilde{r}}{N^{1/2}}\crcr
& + \frac{\tilde{r}}{N}\bigg[6\bar{g}_1\left(1+3\alpha_{S,0}\bar{g}_1\right)+\frac{\tilde{\epsilon}}{2}\left(4\alpha_{D,1}\bar{g}_2+12\alpha_{S,1}\bar{g}^2-1\right)\bigg]+\mathcal{O}(N^{-3/2})\,.
\label{eq:beta-LR}
\end{align}

We then parametrize the critical couplings as:

	\begin{align}
		\bar{g}^{\star} \, &= \,  \, \bar{g}^{\star}_{(0)} \, + \, \bar{g}^{\star}_{(1)} N^{-1/2}  + \, \mathcal{O}(N^{-1}) \, , 
		\crcr
		\bar{g}_{1}^{\star} \, &= \, \bar{g}^{\star}_{1,(0)} \, + \,  \, \bar{g}^{\star}_{1,(1)} N^{-1/2}  + \, \mathcal{O}(N^{-1}) \, , \crcr
		\bar{g}_2^{\star} \, &= \, \bar{g}^{\star}_{2,(0)} \, + \,  \, \bar{g}^{\star}_{2,(1)} N^{-1/2} \, + \, \mathcal{O}(N^{-1}) \, . \label{eq:g-parametrization} 
	\end{align}

\paragraph{Leading-order.}

As we already discussed, at leading order (i.e.\ $N^0$), the tetrahedron beta function is identically zero, hence $\bar{g}^{\star}_{(0)}$ is a free parameter.
For the other two couplings, the leading order fixed points, expanded to second order in $\bar{g}^{\star}_{(0)}$, are:
\begin{align}
\bar{g}^{\star}_{1,(0)}&=\pm \sqrt{-\bar{g}^{\star}_{(0)}{}^2}-\bar{g}^{\star}_{(0)}{}^2\alpha_{S,0}+ \mathcal{O}(\bar{g}^{\star}_{(0)}{}^3)\,, \crcr
\bar{g}^{\star}_{2,(0)}&=\pm \sqrt{3}\sqrt{-\bar{g}^{\star}_{(0)}{}^2} -3\bar{g}^{\star}_{(0)}{}^2\alpha_{S,0}+ \mathcal{O}(\bar{g}^{\star}_{(0)}{}^3)\,.
\label{eq:LO_pillow_dt}
\end{align}
They correspond to the lines of fixed points found at large $N$ in \cite{Benedetti:2019eyl}. For small coupling $|\bar{g}^{\star}_{(0)}|$, $\bar{g}^{\star}_{1,(0)}$ and $\bar{g}^{\star}_{2,(0)}$ are complex for real $\bar{g}^{\star}_{(0)}$ and real for purely imaginary $\bar{g}^{\star}_{(0)}$.

\paragraph{Next-to-leading order.}

Substituting \eqref{eq:g-parametrization} and \eqref{eq:LO_pillow_dt} into the beta functions \eqref{eq:beta-LR} and solving for fixed points at order $N^{-1/2}$ we find $\bar{g}^{\star}_{1,(1)}$ and $\bar{g}^{\star}_{2,(1)}$ in terms of $\bar{g}^{\star}_{(0)}$ and $\bar{g}^{\star}_{(1)}$:

\begin{align}
\bar{g}^{\star}_{1,(1)}&=-2\bar{g}^{\star}_{(0)}-2\bar{g}^{\star}_{(0)}\bar{g}^{\star}_{(1)}\alpha_{S,0}\mp \frac{\bar{g}^{\star}_{(0)}\bar{g}^{\star}_{(1)}}{\sqrt{-\bar{g}^{\star}_{(0)}{}^2}} + \mathcal{O}(\bar{g}^{\star}_{(0)}{}^3)\,, \crcr
\bar{g}^{\star}_{2,(1)}&=-3\bar{g}^{\star}_{(0)}-6\bar{g}^{\star}_{(0)}\bar{g}^{\star}_{(1)}\alpha_{S,0}\mp\frac{\sqrt{3}\bar{g}^{\star}_{(0)}\bar{g}^{\star}_{(1)}}{\sqrt{-\bar{g}^{\star}_{(0)}{}^2}} + \mathcal{O}(\bar{g}^{\star}_{(0)}{}^3) \,.
\end{align}
The signs in the two sets $\{\bar{g}^{\star}_{1,(0)},\bar{g}^{\star}_{1,(1)}\}$ and $\{\bar{g}^{\star}_{2,(0)},\bar{g}^{\star}_{2,(1)}\}$ are taken to be simultaneously either the upper or lower ones so that we still have four choices of sign. 

\paragraph{Fixing the tetrahedron coupling.}

Since the beta function of the tetrahedron is still zero at order $N^{-1/2}$, it would seem that our lines of fixed points have become surfaces (that is parametrized by two free parameters $\bar{g}^{\star}_{ (0)},\bar{g}^{\star}_{ (1)}$). On the other hand, if we homogeneously truncate all the beta functions at this order, there is no real justification for the expansion of $\bar{g}^{\star} $ in \eqref{eq:g-parametrization}; this is only justified at higher orders, as all the orders $N^{-n/2}$ with $n\geq 2$ in the tetrahedron beta function are non-trivial.
In the spirit of a $1/N$ expansion, as opposed to a strict $N\to\infty$ limit, it is more consistent to keep the same number of non-trivial orders for each beta function regardless of their different scaling in $N$. By doing so, we will be able to fix $\bar{g}^{\star}_{(0)}$ and $\bar{g}^{\star}_{(1)}$.

Substituting \eqref{eq:LO_pillow_dt} into the order $N^{-1}$ of the tetrahedron beta function,  we fix $\bar{g}^{\star}_{(0)}$. Besides the trivial solution, we find:
\begin{align}
\bar{g}^{\star}_{(0)}=\pm\frac{1}{2\alpha_{S,0}}\sqrt{2 \pm \frac{ 6+\tilde{\epsilon}\alpha_{S,0}}{\sqrt{3(3+\tilde{\epsilon}\alpha_{S,0})}}}\,.
\label{eq:solg0s}
\end{align}
The choice of signs is independent of the choices for the previous solutions. 

We are interested in purely imaginary solutions, as at leading order this gives real critical exponents \cite{Benedetti:2019eyl}, and a real spectrum of bilinear operators, with real OPE coefficients \cite{Benedetti:2019ikb}. The solutions with a plus sign inside the square root have a non-zero real part for all values of $\tilde{\epsilon}$, in particular remain finite for $\tilde{\epsilon}\to 0$, and thus they are not to be trusted in our perturbative expansion. The solutions with a minus sign instead are purely imaginary for $\tilde{\epsilon}<-3/\alpha_{S,0}$ (notice this bound is positive as $\alpha_{S,0}$ is negative), and they vanish for $\tilde{\epsilon}\to 0$. 

In this case, we can  expand $\bar{g}^{\star}_{(0)} $ for small $\tilde{\epsilon}$, finding:
\begin{equation}
\bar{g}^{\star}_{(0)}=\pm \frac{i }{12}\left(\tilde{\epsilon}-\frac{\alpha_{S,0}}{6}\tilde{\epsilon}^2\right) +\mathcal{O}(\tilde{\epsilon}^3) \,.
\label{eq:gt0_epsilon}
\end{equation}
We can also expand $\bar{g}_{1,(0)}$ and $\bar{g}_{2,(0)}$ in $\tilde{\epsilon}$:
\begin{align}
\bar{g}_{1,(0)}&= \pm \frac{1}{12}\left(\tilde{\epsilon}-\frac{\alpha_{S,0}}{12}(2\mp1)\tilde{\epsilon}^2\right) +\mathcal{O}(\tilde{\epsilon}^3)\,, \crcr
\bar{g}_{2,(0)}&=\pm \frac{1}{4\sqrt{3}}\left(\tilde{\epsilon}-\frac{\alpha_{S,0}}{12}(2\mp\sqrt{3})\tilde{\epsilon}^2\right) +\mathcal{O}(\tilde{\epsilon}^3) \,,
\label{eq:g1-g2_epsilon}
\end{align}
where the global sign and the one inside the brackets are taken to be simultaneously either the upper or lower ones. 

The $N^{-1/2}$ correction $\bar{g}^{\star}_{(1)}$ is still a free parameter at this order. In order to fix it we need to consider the $N^{-3/2}$ contribution to $\beta_t$, which we have not displayed in \eqref{eq:beta-LR}.
This is easily obtained from the general multi-scalar  results of \cite{Benedetti:2020rrq}, from which we find:
\begin{align}
\beta_t \, = & \, \frac{\bar{g}}{N}\Big[12\bar{g_1}\left(1+\alpha_{S,0} \bar{g_1} \right) -\tilde{\epsilon}\Big]+\frac{48}{N^{3/2}}\alpha_{S,0}\bar{g}_1\bar{g}^2 +\mathcal{O}(N^{-2}) \,.
\end{align}
Substituting the coupling $1/N$ expansions from \ref{eq:g-parametrization}, the order $N^{-3/2}$ of $\beta_t $ is:
\begin{equation}
-6\tilde{\epsilon}\bar{g}_{(1)}+72\bar{g}_{1,(1)}\left(\bar{g}_{(0)}+2\bar{g}_{1,(0)}\bar{g}_{(0)}\alpha_{S,0}\right)+72\bar{g}_{1,(0)}\left(\bar{g}_{(1)}+4\bar{g}_{(0)}^2\alpha_{S,0}+\bar{g}_{(1)}\bar{g}_{1,(1)}\alpha_{S,0}\right) \, ,
\end{equation}
and substituting the values of $\bar{g}_{1,(0)}^{\star}$ and $\bar{g}_{1,(1)}^{\star}$, solving for $\bar{g}_{(1)}^{\star}$ in terms of $\bar{g}_{(0)}^{\star}$ we obtain:
\begin{equation}
\bar{g}_{(1)}^{\star}=-\frac{24\bar{g}_{(0)}^{\star}{}^2}{\tilde{\epsilon}\pm\frac{24\bar{g}_{(0)}^{\star}{}^2}{\sqrt{-\bar{g}_{(0)}^{\star}{}^2}}+72\bar{g}_{(0)}^{\star}{}^2\alpha_{S,0}} \,,
\label{eq:gt1_comp}
\end{equation}
where the choice of sign is the same as in $\bar{g}_{1,(0)}^{\star}$.
This expression is real for purely imaginary $\bar{g}_{(0)}^{\star}$. 

The expression \eqref{eq:gt1_comp} comes from a two-loop truncation and thus it should be trusted only up to  order $\tilde{\epsilon}^2$.
Therefore, we first substitute \ref{eq:gt0_epsilon} in \ref{eq:gt1_comp} and then expand at order two in $\tilde{\epsilon}$:
\begin{align}
\bar{g}_{(1)}^{\star}= \begin{cases} \frac{1}{6}\left(-\tilde{\epsilon}+\frac{\alpha_{S,0}}{2}\tilde{\epsilon}^2\right) + \mathcal{O}(\tilde{\epsilon}^3) \;\;\; \text{ for the upper choice of sign,} \\
\frac{1}{18}\left(\tilde{\epsilon}-\frac{\alpha_{S,0}}{18}\tilde{\epsilon}^2\right) + \mathcal{O}(\tilde{\epsilon}^3) \;\;\; \text{ for the lower choice of sign.} 
\end{cases}
\end{align}
We can now also give the $\tilde{\epsilon}$ expansion of $\bar{g}_{1,(1)}^{\star}$ and $\bar{g}_{2,(1)}^{\star}$.
\begin{align}
\bar{g}_{1,(1)}^{\star} &= \begin{cases} \mp \frac{i\alpha_{S,0}}{36}\tilde{\epsilon^2} +\mathcal{O}(\tilde{\epsilon}^3) \;\;\;  \text{ for the upper choice of sign in } \bar{g}_{1,(0)}^{\star}\,, \\
  \mp \frac{i}{9}\left(\tilde{\epsilon}-\frac{5\alpha_{S,0}}{36}\tilde{\epsilon}^2\right)+\mathcal{O}(\tilde{\epsilon}^3)\;\;\; \text{ for the lower choice of sign in } \bar{g}_{1,(0)}^{\star}\,, \end{cases} \crcr
\bar{g}_{2,(1)}^{\star}&= \begin{cases} \pm \frac{i}{12}\left((-3\pm 2\sqrt{3})\tilde{\epsilon}+\frac{\alpha_{S,0}}{2}(3\mp 2\sqrt{3})\tilde{\epsilon}^2\right)+\mathcal{O}(\tilde{\epsilon}^3) \;\;\;  \text{ for the upper choice of sign in } \bar{g}_{1,(0)}^{\star}\,, \\
\pm \frac{i}{36}\left((-9 \mp 2\sqrt{3})\tilde{\epsilon}+\frac{\alpha_{S,0}}{18}(9\pm 2\sqrt{3})\tilde{\epsilon}^2\right)+\mathcal{O}(\tilde{\epsilon}^3) \;\;\;  \text{ for the lower choice of sign in } \bar{g}_{1,(0)}^{\star}\,, \end{cases}
\nonumber
\end{align}
where the choice of sign in front is the same as for $\bar{g}_{(0)}^{\star}$ and the choice of sign in the parenthesis for $\bar{g}_{2,(1)}^{\star}$ is the same as for $\bar{g}_{2,(0)}^{\star}$.

\subsubsection{Critical exponents}

We will now compute the critical exponents. For the quadratic coupling we obtain:
\begin{align}
\partial  \beta^{(2)}(\bar{g}^{\star})&=-\nu^{-1}=-\frac{d}{2} \pm 2\sqrt{-3\bar{g}^{\star}_{(0)}{}^2} \mp \frac{1}{N^{1/2}}\frac{6\bar{g}^{\star}_{(0)}\bar{g}^{\star}_{(1)}}{\sqrt{-3\bar{g}^{\star}_{(0)}{}^2}} +\mathcal{O}(N^{-1},\bar{g}^{\star}_{(0)}{}^3) \,,
\end{align}
where the signs are taken to be simultaneously either the upper or lower ones and are the same as for $\bar{g}_{2,(0)}$.

The critical exponents for the quartic couplings are given by:\footnote{They correspond to the diagonal elements as the stability matrix is triangular at order $\mathcal{O}(N^{-1/2})$.}

\begin{align}
\partial \beta_{1}(\bar{g}^{\star})&=\pm \Bigg[ 4\sqrt{-\bar{g}^{\star}_{(0)}{}^2}-\frac{1}{N^{1/2}}\frac{4\bar{g}^{\star}_{(0)}\bar{g}^{\star}_{(1)}}{\sqrt{-\bar{g}^{\star}_{(0)}{}^2}}\Bigg] +\mathcal{O}(N^{-1},\bar{g}^{\star}_{(0)}{}^3)\,, \crcr
\partial \beta_{2}(\bar{g}^{\star})&=\pm \Bigg[ 4\sqrt{-3\bar{g}^{\star}_{(0)}{}^2}-\frac{1}{N^{1/2}}\frac{12\bar{g}^{\star}_{(0)}\bar{g}^{\star}_{(1)}}{\sqrt{-3\bar{g}^{\star}_{(0)}{}^2}}\Bigg] +\mathcal{O}(N^{-1},\bar{g}^{\star}_{(0)}{}^3) \,,
\end{align}
where the signs are taken to be simultaneously either the upper or lower ones in the two sets $\{\bar{g}_{1,(0)},\partial \beta_{1}\}$ and $\{\bar{g}_{2,(0)},\partial \beta_{2}\}$.
At leading order, the stable fixed points are those with the choice of the upper sign in $\bar{g}_{1,(0)}$ and $\bar{g}_{2,(0)}$. There are two such fixed points depending on the choice of sign in $\bar{g}_{(0)}^{\star}$:
\begin{align}
\bar{g}^{\star}&=\pm \frac{i}{12}\left(\tilde{\epsilon}-\frac{\alpha_{S,0}}{6}\tilde{\epsilon}^2\right)+\frac{1}{6N^{1/2}}\left(\tilde{\epsilon}-\frac{\alpha_{S,0}}{3}\tilde{\epsilon}^2\right)+\mathcal{O}(\tilde{\epsilon}^3,N^{-1}) \,, \crcr 
\bar{g}_{1}^{\star}&=\frac{1}{12}\left(\tilde{\epsilon}-\frac{\alpha_{S,0}}{12}\tilde{\epsilon}^2\right)\mp \frac{i\alpha_{S,0}}{36N^{1/2}}\tilde{\epsilon}^2 +\mathcal{O}(\tilde{\epsilon}^3,N^{-1}) \,, \crcr
\bar{g}_{2}^{\star}&=\frac{1}{4\sqrt{3}}\left(\tilde{\epsilon}-\frac{\alpha_{S,0}}{12}(2-\sqrt{3})\tilde{\epsilon}^2\right)\pm \frac{i(-3+2\sqrt{3})}{12N^{1/2}}\left(\tilde{\epsilon}-\frac{\alpha_{S,0}}{2}\tilde{\epsilon}^2\right)+\mathcal{O}(\tilde{\epsilon}^3,N^{-1}) \,,
\label{eq:stablefpLO}
\end{align}
where the signs in all three couplings are taken to be simultaneously either the upper or lower ones. 
For these two fixed points, the $\tilde{\epsilon}$ expansions of the critical couplings are then: 
\begin{align}
\partial  \beta^{(2)}(\bar{g}^{\star})&=-\nu^{-1}=-\frac{d}{2}+\frac{1}{2\sqrt{3}}\left(\tilde{\epsilon}-\frac{\alpha_{S,0}}{6}\tilde{\epsilon}^2\right)\pm \frac{i}{\sqrt{3}N^{1/2}}\left(\tilde{\epsilon}-\frac{\alpha_{S,0}}{2}\tilde{\epsilon}^2\right)+\mathcal{O}(\tilde{\epsilon}^3,N^{-1}) \,, \crcr
\partial \beta_{1}(\bar{g}^{\star})&=\frac{1}{3}\left(\tilde{\epsilon}-\frac{\alpha_{S,0}}{6}\tilde{\epsilon}^2\right) \pm \frac{2 i }{3N^{1/2}}\left(\tilde{\epsilon}-\frac{\alpha_{S,0}}{2}\tilde{\epsilon}^2\right) +\mathcal{O}(\tilde{\epsilon}^3,N^{-1}) \,, \crcr
\partial \beta_{2}(\bar{g}^{\star})&= \frac{1}{\sqrt{3}}\left(\tilde{\epsilon}-\frac{\alpha_{S,0}}{6}\tilde{\epsilon}^2\right) \pm \frac{2 i }{\sqrt{3} N^{1/2}}\left(\tilde{\epsilon}-\frac{\alpha_{S,0}}{2}\tilde{\epsilon}^2\right) +\mathcal{O}(\tilde{\epsilon}^3,N^{-1}) \,,
\end{align}
where the choice of sign is the same as in $\bar{g}_{(0)}^{\star}$.

In order to compute the critical exponent of the tetrahedron coupling, we need to compute the eigenvalues of the stability matrix as it is not triangular beyond order
$N^{-1/2}$. However, up to order $N^{-3/2}$, it depends only on the values of the critical couplings at leading and next-to-leading order. For the fixed point in \eqref{eq:stablefpLO}, we have at second order in $\tilde{\epsilon}$:
\begin{equation}
\omega_{t}=\frac{\tilde{\epsilon}}{N}\left(1+\frac{\alpha_{S,0}}{6}\tilde{\epsilon}\right)+\frac{2 i \alpha_{S,0}\tilde{\epsilon}^2}{3N^{3/2}}+\mathcal{O}(\tilde{\epsilon}^3, N^{-2}) \,.
\end{equation}

\medskip

In summary, while at leading order an imaginary tetrahedron coupling leads to four stable fixed lines of real pillow and double-trace couplings, going up to next-to-leading non-trivial order for all the beta functions fixes all the couplings to eight isolated fixed points, having the same reality properties as before at leading order, but the opposite one at subleading order (i.e.\ real tetrahedron and purely imaginary pillow and double-trace corrections). As with the fixed point values, we have for the critical exponents that what was real at leading order gets an imaginary part at subleading order. 

\section{Conclusions}
\label{sec:concl}

We have studied a tri-fundamental model, that is, a multi-scalar model invariant under $O(N_1)\times O(N_2) \times O(N_3)$ transformations, of which the scalar fields form a tri-fundamental representation. We have considered versions of the model with either short or long-range Gaussian part, and we have studied the renormalization group beta functions at finite or large $N_i$, in various scaling limits.
Our main conclusion is that in general we find \emph{no} stable real fixed points with non-zero tetrahedral coupling.

In order to find genuine infrared-stable fixed points with non-zero tetrahedral coupling we have to consider complex fixed points. This immediately raises the prospect that the fixed point theories are not unitary; however, complex CFTs have been considered in statistical physics and in the description of walking behavior in high-energy physics (see for example \cite{Gorbenko:2018ncu,Gorbenko:2018dtm} and references therein). Complex, stable (in all directions) infrared fixed points are obtained in the homogeneous (i.e.\ $N_i=N$ for $i=1,2,3$) large-$N$ limit
of the long-range model. In this case the tetrahedral coupling is exactly marginal, and when taken to be purely imaginary all the CFT data available to us indicates that the leading large-$N$ CFT is real and within unitarity bounds \cite{Benedetti:2019eyl,Benedetti:2019ikb,Benedetti:2020yvb}. In this paper we have shown that this does not survive at subleading order in $1/N$: the line of fixed point reduces to an isolated point, and unitarity is broken by the $1/N$ corrections which bring imaginary parts to the critical exponents.
A similar complex CFT, providing subleading corrections to that of \cite{Giombi:2017dtl}, is found also for the short-range model, but in that case it is the real part of the critical exponents which is suppressed in $1/N$, rather than the imaginary part; therefore, while the two models have probably qualitatively similar behavior at finite $N$, it is only in the long-range case that a real and unitary CFT arises in the strict large-$N$ limit.

A subtle aspect of our analysis of subleading corrections in $1/N$ to the fixed points of the $O(N)^3$ model is the identification of an appropriate hierarchy between the two small parameters at play, i.e.\ $1/N$ and $\epsilon$, the latter being defined as the deviation from the critical dimension in the short-range case, i.e.\ $\epsilon=4-d$, or as the deviation from the critical scaling of the propagator in the long-range case, i.e.\  $C(p)=1/p^{(d+\epsilon)/2}$.
In the former case it turns out that we need $\epsilon N^2 \gg 1$, while in the latter we need $\epsilon N \ll 1$.
The reason for that is the form of the tetrahedron beta functions, which we can roughly understand in the following way. Slightly simplifying things (in reality we have a coupled system of equations), at two-loop order the tetrahedron beta function has the form $\b_{SR}(g)=-\epsilon g + b g^3 +\frac{a}{N} g^2 +\cO(N^{-3/2})$ in the short-range case, and $\b_{LR}(g)= -\epsilon g + \frac{a}{N} g^2 +\cO(N^{-3/2})$ in the long-range case, for some constants $a$ and $b$ of order one.
The conditions on $\epsilon$ and $N$ then arise from demanding that the fixed point from the leading order in $1/N$ remains dominant in the beta function. As a perturbative solution of $\b_{SR}(g^\star)=0$ at leading order implies $g^\star\sim \sqrt{\epsilon}$, we see that the first two terms in $\b_{SR}(g)$ are of order $\epsilon^{3/2}$, while the third is of order $\epsilon/N$, hence we must have $\sqrt{\epsilon}\gg 1/N$.
For the long-range case, a non-trivial perturbative solution of $\b_{LR}(g^\star)=0$ at leading order is instead not possible for $\epsilon>0$, and we must require $\epsilon\ll 1/N$, so that the first two terms in $\b_{LR}(g)$ lead to a Wilson-Fisher type solution, with $\epsilon N$ being the effective small parameter.
A similar tuning of $\epsilon$ and $N$ as in our long-range model was also considered in \cite{Fleming:2020qqx} in order to find a finite-$N$ precursor of the line of fixed points that appear in the short-range $O(N)$ model with $(\phi^2)^3$ interaction at  large-$N$, for $\epsilon=0$.

It would be interesting to understand if the non-existence of stable real fixed points with non-vanishing tetrahedral coupling could be proved in general terms, for example by using group-theoretical arguments, as in \cite{Michel:1985,Toledano:1985}, or by exploiting the gradient flow representation of the renormalization group equations, along the lines of other proofs, for example as in  \cite{Michel:1983in,ZinnJustin:2007zz,Rychkov:2018vya}. We have tried the second route, but failed so far in this task; nonetheless, we report in appendix \ref{app:grad_flow} some relevant formulas for the gradient flow of the tri-fundamental model, hoping that they could serve as reference or inspiration for a future proof.

More generally, it would also be interesting to understand whether any stable real fixed points exist with rank-$p$ tensor symmetry, such as $O(N)^p$, for higher $p$.
We notice also that real fixed points have been found in short-range models with $p=3$, but with sextic interaction,  for small $\epsilon=3-d$\cite{Giombi:2018qgp,Benedetti:2019rja}; it would be interesting to understand if they also become complex at subleading orders, or whether sextic interactions have some fundamental difference with respect to quartic ones.

\section*{Acknowledgements}

We thank Kenta Suzuki for collaboration at the early stages of this work, and for useful comments at a later stage.
This work is supported by the European Research Council (ERC) under the European Union's Horizon 2020 research and innovation program (grant agreement No818066). The work of RG is also supported by Deutsche Forschungsgemeinschaft (DFG, German Research Foundation) under Germany's Excellence Strategy EXC-2181/1 - 390900948 (the Heidelberg STRUCTURES Cluster of Excellence) and
partly supported by Perimeter Institute for Theoretical Physics.

\clearpage

\appendix

\section{Gradient flow} 
\label{app:grad_flow}

We wish to write the beta functions \eqref{eq:beta^(4)} as a gradient flow:
\begin{equation}
\beta_{a}=T_{ab}\frac{\partial U}{\partial g_b}
\label{eq:gradient_flow}
\end{equation}
with $U$ a potential, and $T_{ab}$ a non-trivial symmetric matrix, where the indices $a,b$ run over the five couplings $t$, $p_i$, $d$. For the general system \eqref{eq:beta-general}, the one loop potential is:
\begin{equation}
U_{MS}=-\frac{\epsilon}{2}\tilde{g}_{ijkl}\tilde{g}_{ijkl}+\tilde{g}_{ijkl}\tilde{g}_{klmn}\tilde{g}_{mnij} \,,
\end{equation}
and we recall that the metric in the general case is trivial at this order (in fact it is trivial up to two loops \cite{Wallace:1974dy}). Substituting \eqref{eq:coupling-trifund} we find the one loop potential for the short-range tri-fundamental model:
\begin{align}
U&=-3\epsilon N_1N_2N_3\Bigg[\left(N_1N_2N_3+N_1+N_2+N_3+2\right)\tilde{g}^2+2\left(2+N_1N_2N_3\right)\tilde{g}_d^2 +4(N_1+N_2+N_3)\tilde{g}\tilde{g}_d \crcr
&\quad +\sum_{i=1}^3\Bigg(\left((1+N_i)N_{i+1}N_{i+2}+N_i+3\right)\tilde{g}_{p,i}^2+2\tilde{g}\left(2+(1+N_i)(N_{i+1}+N_{i+2})\right)\tilde{g}_{p,i}\crcr
& \quad \qquad \quad +2\left((1+N_i)(1+N_{i+1})+2N_{i+2}\right)\tilde{g}_{p,i}\tilde{g}_{p,i+1}+4\tilde{g}_d\left(1+N_i+N_{i+1}N_{i+2}\right)\tilde{g}_{p,i}\Bigg) \Bigg] \crcr
& +4N_1N_2N_3\Bigg[\Big(12+\sum_{i=1}^3\left(6N_i+6N_iN_{i+1}+N_i^2(N_{i+1}+N_{i+2})\Big)\right)\tilde{g}^3+2\left(N_1N_2N_3+8\right)\left(N_1N_2N_3+2\right)\tilde{g}_d^3\crcr
& \quad +6\Big(3N_1N_2N_3+\sum_{i=1}^3\left(N_i^2+2N_iN_{i+1}+3N_i\right)+6\Big)\left(\tilde{g}^2\tilde{g}_d+2\tilde{g}_{p,1}\tilde{g}_{p,2}\tilde{g}_{p,3}\right)\crcr
& \quad +6\left(N_1+N_2+N_3\right)\left(N_1N_2N_3+8\right)\tilde{g}_d^2\tilde{g} \crcr
& \quad +\sum_{i=1}^3 3\tilde{g}_{p,i}\Bigg(\left(N_{i+1}^2N_{i+2}^2(1+N_i)+N_{i+1}N_{i+2}(N_i^2+6N_i+13)+N_i^2+13N_i+18\right)\frac{\tilde{g}_{p,i}^2}{3}\crcr
& \qquad  +\left((N_{i+1}^2+N_{i+2}^2)(1+N_i)+N_{i+1}N_{i+2}(N_i^2+3N_i+6)\right. \crcr
& \hspace{2cm} \left. +(N_{i+1}+N_{i+2})(3N_i+7)+N_i^2+9N_i+10\right)\tilde{g}^2 \crcr
& \qquad  + 2\left(8+N_1N_2N_3\right)\left(1+N_i+N_{i+1}N_{i+2}\right)\tilde{g}_d^2 \crcr
& \qquad  +\left(2N_{i+1}N_{i+2}+(N_i+1)(N_{i+1}N_{i+2}(N_{i+1}+N_{i+2})+8)+(N_{i+1}+N_{i+2})(N_i^2+5N_i+10)\right)\tilde{g}\tilde{g}_{p,i}\crcr
& \qquad +2\left(N_i^2+5N_i(1+N_{i+1}N_{i+2})+N_{i+1}N_{i+2}(5+N_{i+1}N_{i+2})+10\right)\tilde{g}_d\tilde{g}_{p,i} \crcr
&\qquad  +\sum_{j,k=1 ; j\neq k \neq i}^3\left(2N_jN_k^2+(1+N_j)(N_i^2+5N_i+10)+N_k(1+N_i)(N_j^2+N_j+8)\right)\tilde{g}_{p,j}\tilde{g}_{p,i} \crcr
& \qquad  + 2\left(2N_{i+2}^2+2N_{i+2}(2N_iN_{i+1}+3(N_i+N_{i+1})+2)+N_i^2+N_{i+1}^2 \right. \crcr
& \hspace{2cm} \left. +(N_i+N_{i+1})(N_iN_{i+1}+7)+2N_{i}N_{i+1}+12\right)\tilde{g}\tilde{g}_{p,i+1} \crcr
& \qquad  +4\left(N_{i+2}^2N_iN_{i+1}+N_{i+2}(N_i^2+N_{i+1}^2+N_i+N_{i+1}+6)+4(1+N_i)(1+N_{i+1})\right)\tilde{g}_d\tilde{g}_{p,i+1} \crcr
& \qquad  +4\left(N_i^2+N_i(N_{i+1}N_{i+2}+4(N_{i+1}+N_{i+2})+1)+(N_{i+1}+N_{i+2})(N_{i+1}N_{i+2}+4)+6\right)\tilde{g}\tilde{g}_d  \Bigg) \Bigg] \,,
\end{align}
where $i \in \{1,2,3\}$ and $N_4=N_1$, $N_5=N_2$.

The matrix $T$ can be now found following \cite{Wallace:1974dy}, by using the following expression for its inverse:
\be
(T^{-1})_{ab} = \f{\p \gt_{\mba \mbb \mbc \mbd} }{\p g_a} \f{\p \gt_{\mba \mbb \mbc \mbd} }{\p g_b} \,.
\ee
Defining $(6 N_1 N_2 N_3)\, \eta_{ab}=(T^{-1})_{ab}$, we have:
\begin{align}
\eta_{tt}&= 2+N_1+N_2+N_3+ N_1 N_2 N_3 \,,\crcr
\eta_{p_i p_i}&=2+(1+N_i)(1+N_{i+1}N_{i+2}) \,,\crcr
\eta_{d d}&=4+2 N_1N_2N_3 \,,\crcr
\eta_{t p_i}&=\eta_{p_i t}=2+(1+N_i)(N_{i+1} + N_{i+2}) \,,\crcr
\eta_{t d}&=\eta_{d t}=2(N_1+N_2+N_3) \,,\crcr
\eta_{p_i p_j}&=1+N_i+N_j+N_iN_j+2N_k  \,, \;\; \text{ with } i\neq j \neq k \in \{1,2,3\} \,, \crcr
\eta_{p_id}&=\eta_{d p_i}=2(1+N_i+N_{i+1}N_{i+2}) \,.
\end{align}

The long-range beta functions \eqref{eq:beta_abcd_alpha}, or \eqref{eq:beta-LR} differ from the short rang ones only by the presence of the $\a_D$ and $\a_S$ coefficients and the absence of the terms coming from the wave-function renormalization. In particular they are equal at one loop.

In the homogeneous large-$N$ limit $N_1=N_2=N_3=N$ we switch to rescaled variables, as in \eqref{eq:bar_g}. Accordingly, the system becomes
\be \label{eq:gradient_flow_bar}
 6 N^3 \bar{\eta}_{ab} \bar{\beta}_b = \frac{\partial U}{\partial \bar{g}_a} \,,
\qquad 
\bar{\eta}_{ab} = \frac{\p g_c}{\p \bar{g}_a} \eta_{cd} \frac{\p g_d}{\p \bar{g}_b} \, .
\ee
In order to obtain a finite limit, it turns out one needs to first multiply both sides of \eqref{eq:gradient_flow_bar} by
the diagonal matrix $\rho={\rm diag}(N^{-3},N^{-2},N^{-2},N^{-2},1)$ to obtain:
\be
\lim_{N\to\infty} N^3\, \rho \, \bar{\eta} = 
\begin{pmatrix}    
1 & 0 & 0 & 0 & 0 \\
0 & 1 & 0 & 0 & 0 \\
0 & 0 & 1 & 0 & 0 \\
0 & 0 & 0 & 1 & 0 \\
0 & 2 & 2 & 2 & 2 
\end{pmatrix} \,.
\ee
The mixing elements between pillows and double-trace are explained by the diagonalization of the system at large $N$ in Eq.~\eqref{eq:g_1,2}.

Notice that $U$ by itself does not have a finite limit for $N\to\infty$ even when written in terms of $\bar{g}_a$ couplings, it is only $\rho_{ab} \partial U/\partial \bar{g}_b$ that does. However, there is no need to rescale by $\rho$ if we write the system as in \eqref{eq:gradient_flow}.

\bibliographystyle{JHEP}
\bibliography{Refs-TMV,Refs-CFT,Refs-QFT} 

\addcontentsline{toc}{section}{References}


\end{document}